\documentclass[reprint,twocolumn,superscriptaddress, amssymb,aps,prl]{revtex4-1}

\usepackage{graphicx}
\usepackage{xcolor} 
\usepackage{amsmath}
\usepackage{hyperref}

\begin{document}
\title{Bichromatic Imaging of Single Molecules in an Optical Tweezer Array}
\author{Connor M. Holland}
\affiliation{Department of Physics, Princeton University, Princeton, New Jersey 08544 USA}
\author{Yukai Lu}
\affiliation{Department of Physics, Princeton University, Princeton, New Jersey 08544 USA}
\affiliation{Department of Electrical and Computer Engineering, Princeton University, Princeton, New Jersey 08544 USA}
\author{Lawrence W. Cheuk}
\email{lcheuk@princeton.edu}
\affiliation{Department of Physics, Princeton University, Princeton, New Jersey 08544 USA}

\date{\today}
\begin{abstract}

We report on a novel bichromatic fluorescent imaging scheme for background-free detection of single CaF molecules trapped in an optical tweezer array. By collecting fluorescence on one optical transition while using another for laser-cooling, we achieve an imaging fidelity of $97.7(2)\%$ and a non-destructive detection fidelity of $95.5(6)\%$. We characterize loss mechanisms of our scheme, many of which are generically relevant to the fluorescent detection of trapped molecules, including two-photon decay and admixtures of higher excited states that are induced by the trapping light. \end{abstract}
\maketitle

Single molecules trapped in rearrangeable arrays of optical tweezers have been proposed as a platform for quantum simulation, quantum information processing, and precision measurements~\cite{Ni2018gate,Blackmore2018QuantumReview, Yu2019symtop, Mitra2022MolfundphysReview}. Compared to the platform of neutral atoms in optical tweezer arrays ~\cite{Endres2016atomarray,Labuhn2016atomarray,Cooper2018AEA,Norcia2018AEA,Saskin2019AEA}, which has been very successful in a wide variety of quantum applications~\cite{Norcia2019clock,Graham20192Dgate,Levine2019gate,Madjarov2020entangle,Ma2022YbGate,Bernien201751atom,Leseleuc2019SSH}, molecules in tweezer arrays offer new capabilities thanks to their rich internal structure and long-range electric dipolar interactions in long-lived states~\cite{Carr2009review,Bohn2017molreview}. These features have spurred experimental efforts to create single molecules in tweezer arrays via coherent assembly from constituent atoms~\cite{Ni2008groundstate,Liu2019NaCsTweezer,Zhang2020NaCsMagAssoc,Zhang2021NaCsArray,Spence2022rbcs} and through direct laser-cooling~\cite{Anderegg2019Tweezer,Cheuk2020collisions,burchesky2021rotcoh}. The latter approach of directly laser-cooling has the advantage of potentially higher loading rates and higher detection fidelities, because individual molecules can be directly cooled and read out optically through cycling hundreds of photons. Crucially, it appears to be a general technique that could be extended to polyatomic molecules~\cite{kozyryev2017sisyphus,Augenbraun2020asymmcooling,mitra2020CaOCH3,baum2021CaOH,Vilas2022MOTCaOH}. 

Laser-cooled molecules in optical tweezers may be non-destructively detected through fluorescent imaging. However, because optical tweezers have limited trap depths, cooling is needed to counteract the recoil heating due to photon scattering. To date, $\Lambda$-imaging has been the only technique that provides both cooling and the fluorescence needed to detect single molecules~\cite{Cheuk2018Lambda,Anderegg2019Tweezer}. Nevertheless, it comes with a side effect. Stray laser-cooling light can lead to significant detection backgrounds that reduce detection fidelities. A strategy to circumvent this problem is to use two optical transitions, one for laser-cooling and the other for inducing fluorescence. In such a bichromatic scheme, background-free detection can be achieved by spectrally filtering out stray laser-cooling light. This approach has been successfully used to image single atoms in optical lattices and optical tweezers~\cite{Haller2015EIT, Cheuk2015qgm, Yamamoto2016YbQGM, Cooper2018AEA, Norcia2018AEA,Saskin2019AEA}. For molecules, bichromatic imaging schemes have also been demonstrated for laser-cooled ensembles~\cite{Vilas2022MOTCaOH,Shaw2021ramancycle}, but not with single molecule resolution.

\begin{figure}[!]
	{\includegraphics[width=\columnwidth]{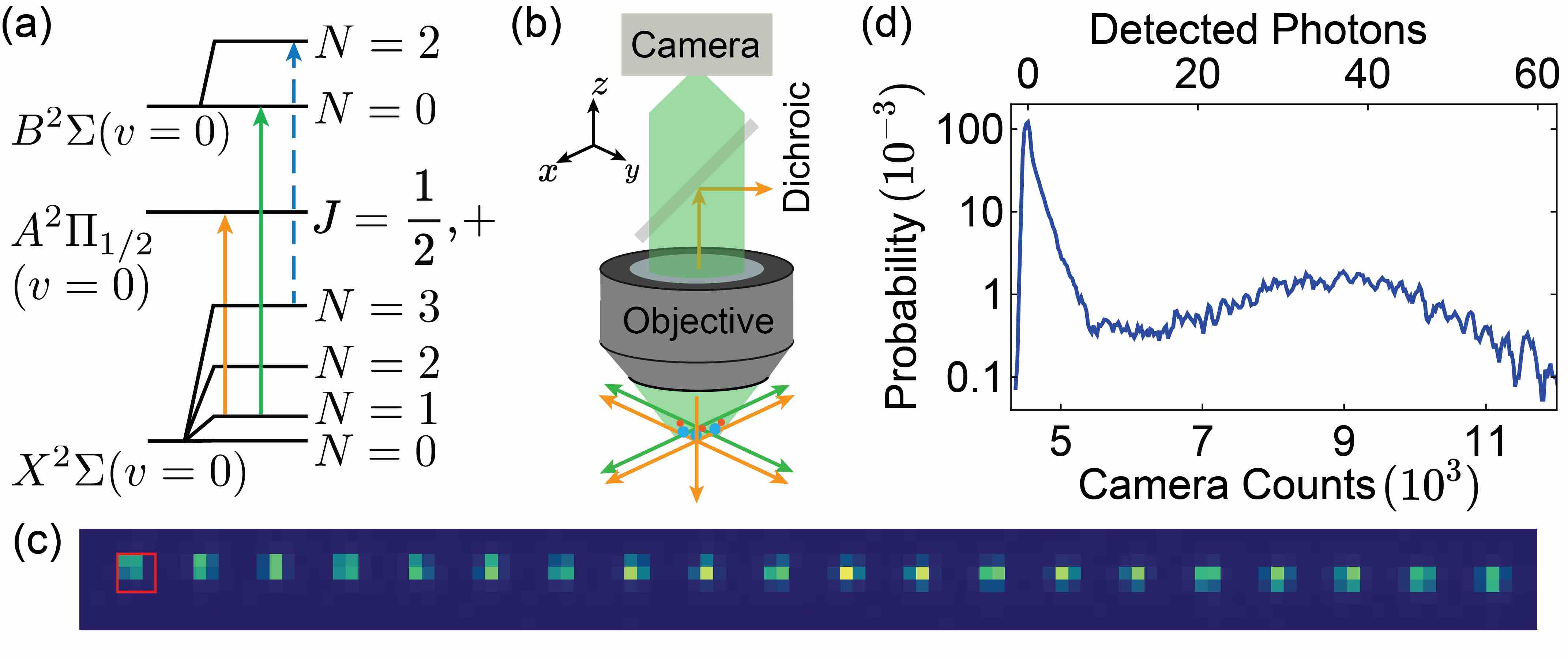}}
	\caption{\label{fig_1} (a) Relevant CaF levels and Addressed Transitions for Bichromatic imaging. Shown in orange (green) is the $X-A$ $\Lambda$-cooling light  ($X-B$ imgaging light) at 606\,nm (531\,nm). Also indicated in blue is a rotational repumper addressing the $X(v=0,N=3) - B(v=0,N=2)$ transition. (b)  $\Lambda$-cooling light (orange) illuminates molecules along three directions ($\pm \hat{x}, \pm \hat{y},\pm \hat{z}$) while $X-B$ imaging light is sent approximately along the $\pm \hat{x}$ and $\pm \hat{y}$ directions. $X-B$ fluorescence (green) enters the objective, passes through dichroic filters, and is detected by an EMCCD camera. (c) Average image of molecules in a 20-site optical tweezer array. The red square indicates an exemplary cropped region used to obtain imaging histograms. (d) Example histogram at an average tweezer occupation of $p=0.21$.}
	\vspace{-0.2in}
\end{figure}

In this paper, we demonstrate a high-fidelity bichromatic imaging scheme for single CaF molecules trapped in optical tweezers. Our work starts with laser-cooled CaF loaded from a magneto-optical trap~\cite{Truppe2017Mot,Anderegg2017MOT} into a 1D optical lattice formed by a retro-reflected 1064\,nm laser beam~\cite{Lu2021Ring}. The molecules are transported to the focus of a high numerical aperture microscope objective (NA=0.65) and then loaded in the presence of $\Lambda$-cooling light into a linear array of 20 identical optical tweezer traps~\cite{supplement}. The tweezer traps are created from focused beams of 781\,nm light propagating through the objective along $-\hat{z}$ (Fig.~\ref{fig_1}(b)). Each tweezer trap has a Gaussian beam waist of $w_0=720(14)\,\text{nm}$, and has a trap depth of $V = k_B \times 1.28(11)\,\text{mK}$ for molecules in the electronic ground state $X^2\Sigma(v=0)$. 
 
In our bichromatic imaging scheme, we apply $\Lambda$-cooling light on the $X^2\Sigma(v=0,N=1) \rightarrow A^2\Pi_{1/2}(v=0,J=1/2,+)$ transition at 606\,nm, along with light addressing the $X^2\Sigma(v=0,N=1,F=0) \rightarrow B^2\Sigma(v=0, N=0)$ transition at 531\,nm. Vibrational repumpers addressing the $X^2\Sigma (v=1,2,3,N=1) \rightarrow A^2\Pi_{1/2}(v=0,1,2, J=1/2,+)$ transitions are present. Additionally, light addressing the $X^2\Sigma(v=0,N=3) \rightarrow B^2\Sigma(v=0, N=2)$ transition is applied to repump molecules that decay into $X^2\Sigma(v=0,N=3)$  (Fig.~\ref{fig_1}(a)). The $\Lambda$-cooling light is applied along all three directions ($\pm\hat{x}$, $\pm\hat{y}$ and $\pm\hat{z}$), while the $X-B$ excitation light avoids the objective axis ($\hat{z}$) and only propagates roughly in the $\hat{x}-\hat{y}$ plane. The resulting $X-B$ fluorescence is collected through the microscope objective and imaged onto an EMCCD camera with an overall efficiency of 0.049(2). Dichroic filters in the imaging path remove the $\Lambda$-cooling light, $X-A$ fluorescence, and stray optical tweezer light. As shown in Fig.~\ref{fig_1}(c), the average images obtained using bichromatic imaging reveal the presence of molecules in the optical tweezer array. Optimizing for maximal survival at a fixed $X-B$ light intensity of $I=12(1)\text{mW/cm}^2$, we find $\Lambda$-cooling parameters similar to those previously reported for $\Lambda$-imaging~\cite{Anderegg2019Tweezer}. At optimal parameters, we obtain an imaging lifetime of $\tau = 148(14)\,\text{ms}$, and a $X-B$ photon scattering rate of $\Gamma_{\text{sc}}=1.95(4)\times10^4\,\text{s}^{-1}$. 

\begin{figure}
	{\includegraphics[width=\columnwidth]{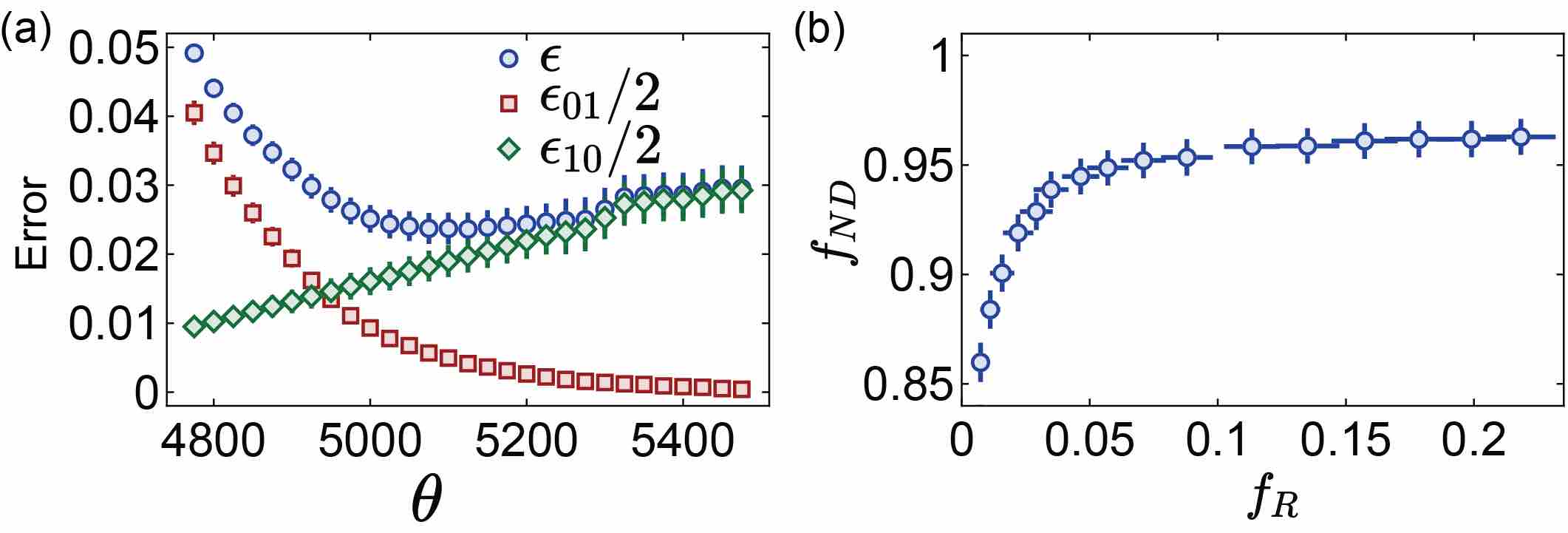}}
	\caption{\label{fig_2} (a) Error probabilities versus classification threshold $\theta$. Shown in green diamonds and red squares are the measured error probabilities $\epsilon_{10}/2$ and $\epsilon_{01}/2$, respectively. The total error $\epsilon$ for an average occupation of $p=0.50$ is shown in blue circles. $\epsilon$ is minimal at an optimal threshold of $\theta_{\text{opt}} \approx 5100$ and corresponds to an imaging fidelity of $f=0.977(2)$. (b) Shown in blue circles is the non-destructive fidelity $f_{\text{ND}}$ versus the data rejection rate $f_\text{R}$. Both $f_{\text{ND}}$ and $f_\text{R}$ are controlled by the first image classification threshold $\theta$. By increasing $\theta$, the fidelity $f_\text{ND}$ can be improved at the expense of a higher data rejection rate $f_\text{R}$.}
		\vspace{-0.2in}
\end{figure}

To evaluate the imaging performance, we first construct histograms of the total camera counts in cropped image regions surrounding each tweezer (Fig.~\ref{fig_1}(c)). The histograms reveal a bimodal distribution (Fig.~\ref{fig_1}(d)), with the two peaks corresponding to tweezer instances that are empty and occupied.
We classify tweezer instances below (above) a threshold $\theta$ as empty (occupied). We describe the classification errors by two probabilities, $\epsilon_{10}$ and $\epsilon_{01}$, which correspond to the probabilities of incorrectly classifying an occupied tweezer as empty and vice versa. The total misclassification probability is dependent on the average occupation $p$ and is given by $\epsilon(p)=p\, \epsilon_{10} + (1-p)\epsilon_{01}$. We define the imaging fidelity $f$ as probability of correctly classifying a tweezer when the occupation probability is 0.5, i.e. $f=1-\epsilon_{\text{opt}}(0.5)$, at the optimal threshold $\theta_{\text{opt}}$. By applying a rescaling procedure to histograms (details in~\cite{supplement}), we obtain the classification errors $\epsilon_{10}$ and $\epsilon_{01}$ for 30\,ms-long images, from which we extract an imaging fidelity of $f=0.977(2)$ (Fig.~\ref{fig_2}(a)). During the images, $\approx 35$ photons are collected per molecule, which is well above the background of $\approx 0.3$ photons per tweezer.

A second metric for single molecule detection is the non-destructive detection fidelity $f_{\text{ND}}$, which we define as the probability that a tweezer initially classified as occupied remains occupied following imaging. This metric is relevant to rearranging molecules into defect-free arrays, since the success probability of creating an array of size $N$ is limited by $\left(f_{\text{ND}}\right)^N$. Unlike the imaging fidelity $f$, the non-destructive detection fidelity $f_{\text{ND}}$ primarily depends on $\epsilon_{01}$, which is approximately equivalent to the false-positive rate. The detection threshold $\theta$ can therefore be increased to reduce $\epsilon_{01}$ at the expense of increasing $\epsilon_{10}$, which is the probability of rejecting occupied tweezers. In addition to the classification errors $\epsilon_{10}$ and $\epsilon_{01}$, $f_{\text{ND}}$ is also affected by the survival probability of molecules following detection. Thus, the imaging duration should be chosen to balance between minimizing imaging loss and maximizing molecular fluorescence; we find that a 10\,ms-long imaging duration strikes this balance. Since $f_{\text{ND}}$ is the fraction of molecules initially classified as occupied that survives detection, it can be directly measured with repeated imaging. To determine $f_{\text{ND}}$, we image the molecules twice, first with a 10\,ms-long bichromatic pulse, and then with a  30\,ms-long pulse~\cite{supplement}. As expected, $f_{\text{ND}}$ increases at the expense of rejecting occupied tweezer instances (Fig.~\ref{fig_2}(b)). At a moderate data rejection rate of $f_{\text{R}}=15\%$, we achieve a non-destructive detection fidelity of $f_{\text{ND}}=95.5(6)\%$. Both the imaging fidelity and non-destructive detection fidelity are comparable to or better than previously reported~\cite{Anderegg2019Tweezer,Cheuk2020collisions}, and are sufficient for experiments with small-scale rearrangeable optical tweezer arrays.

\begin{figure}
	{\includegraphics[width=\columnwidth]{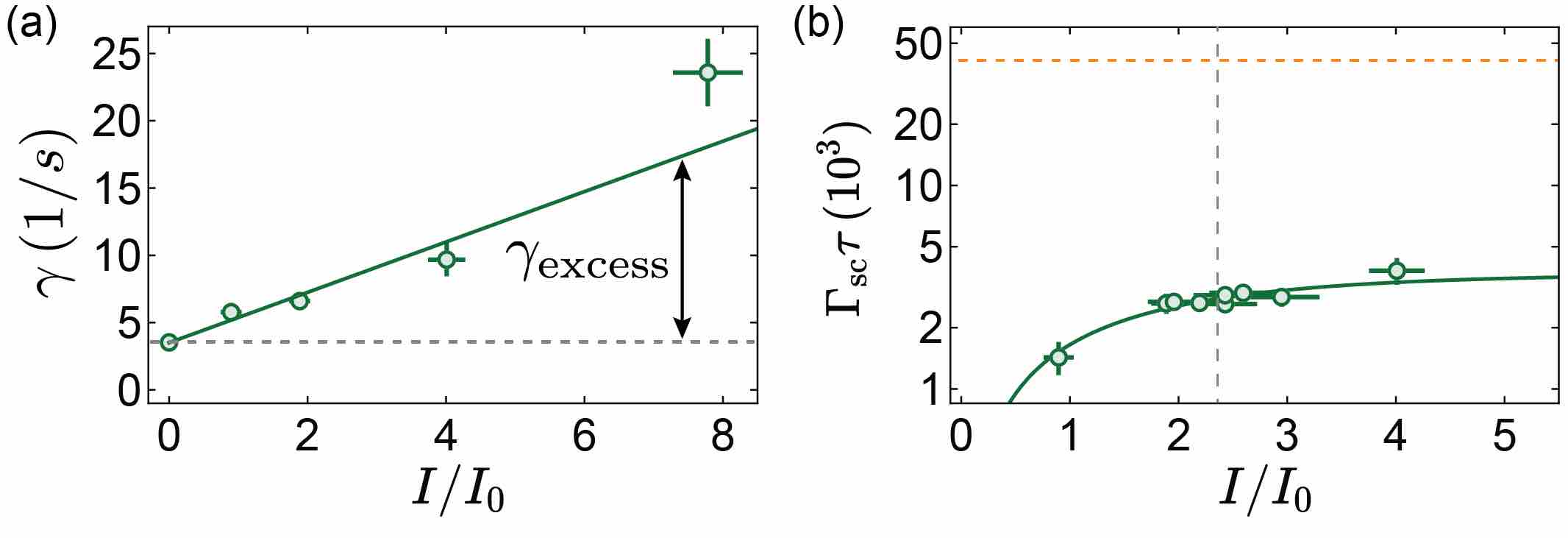}}
	\caption{\label{fig_3} (a) Green circles show the bichromatic imaging loss rate $\gamma$ versus the normalized $X-B$ imaging intensity $I/I_0$ ($I_0 = 5.2(5)\,\text{mW/cm}^2$). The solid line shows a linear fit. (b) The figure-of-merit $\eta = \Gamma_{\text{sc}}\tau$ for bichromatic imaging (green circles) saturates to $\eta\approx 3\times10^3$ at high intensities, far lower than that for single-color $\Lambda$-imaging (orange dashed line, $\eta \approx 40\times 10^3$). The green curve shows the fit to an exponential saturated curve. The vertical dashed line indicates the $X-B$ intensity used for bichromatic imaging in all measurements. At this intensity, $\eta= 2.7(2)\times10^3$. }
	\vspace{-0.2in}
\end{figure}

Having characterized bichromatic imaging fidelities, we next compare the bichromatic imaging performance to that of single-color $\Lambda$-imaging. We first examine the dependence of the bichromatic loss rate $\gamma = 1/\tau$ on the $X-B$ imaging power $I$ and find that it increases with $I$. In particular, the excess loss rate beyond $\Lambda$-imaging loss, given by $\gamma_{\text{excess}} = \gamma(I)-\gamma(0)$, is observed to increase linearly with $I$ (Fig.~\ref{fig_3}(a)). To allow further comparison with single-color $\Lambda$-imaging, we define an imaging figure-of-merit $\eta = \Gamma_{\text{sc}} \tau$, where $\Gamma_{\text{sc}}$ is the fluorescence rate of imaging photons and $\tau$ is the imaging lifetime. We note that $\eta$ measures how many photons can be collected before a molecule is lost. At high $X-B$ powers, $\eta$ saturates to $\approx 3\times10^3$, roughly 10 times lower than that of $\Lambda$-imaging ($\eta\approx 40\times10^3$) (Fig.~\ref{fig_3}(b)). Each scattered $X-B$ photon during bichromatic imaging gives $\approx 10$ times more loss compared to each $X-A$ photon scattered during $\Lambda$-imaging.

\begin{figure}[h!]
	\centering
	{\includegraphics[width=\columnwidth]{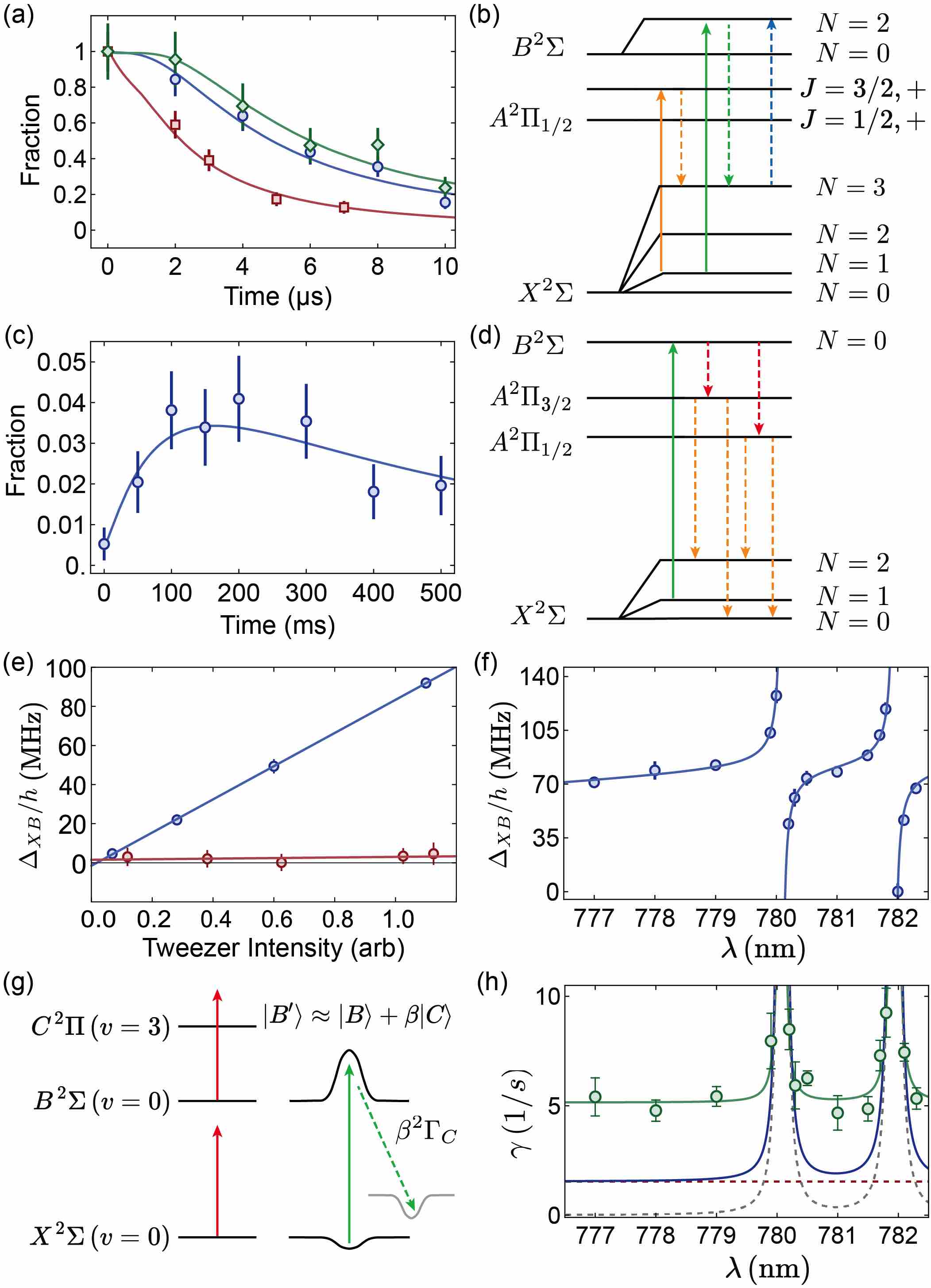}}
	\caption{\label{fig_4} (a) Normalized release-and-recapture fraction versus expansion time. Shown in green diamonds, blue circles, and red squares are release-and-recapture curves after 100\,ms of bichromatic imaging, no imaging, and 10\,ms of $X-B$ excitation alone, respectively. Solid curves are the corresponding fits to Monte-Carlo simulations. (b) Parity-conserving rotational loss due to off-resonant scattering. The relevant transitions excited off-resonantly are shown by the solid orange and green arrows. Decay channels are shown by the dashed orange and green arrows. The blue arrow indicates the $N=3$ rotational repumper. (c) Detected $X^2\Sigma(v=0,N=0,F=1)$ fraction versus bichromatic imaging duration at $I=4.8(4)I_0$. Solid curve shows the fit to a rate equation model~\cite{supplement}. (d) Parity-changing rotational loss due to two-photon $B\rightarrow A \rightarrow X$ decay.  (e) Maximum differential ac Stark shift $\Delta_{XB}$ versus tweezer intensity, normalized to the tweezer imaging intensity. (f) $\Delta_{XB}$ versus tweezer wavelength $\lambda$. (g) Decay induced by $C^2\Pi$-state admixture into the $B^2\Sigma(v=0)$-state. Tweezer light (red) gives rise to repulsive trapping for the $B^2\Sigma(v=0)$ state by admixing in the near-resonant $C^2\Pi$ states. The dressed state $\left | B'\right \rangle$ decays into dark rovibrational states (gray) due to $C$-state admixture. (h) Show in green circles is the measured excess loss rate $\gamma_{\text{excess}}$ versus trapping wavelength $\lambda$, at $X-B$ intensity of $I=2.4(2)I_0$. The red (gray) dashed curve shows the estimated parity-changing loss ($C$-state admixture loss); the solid blue curve is the total estimated excess loss due to both contributions.}
	\vspace{-0.2in}
\end{figure}

We next explore possible mechanisms that could explain the excess loss rate $\gamma_{\text{excess}}$ observed during bichromatic imaging. We first examine whether the excess loss can be explained by higher molecular temperatures. We perform release-and-recapture thermometry after $100\,\text{ms}$ of $\Lambda$-cooling and bichromatic imaging, respectively, and obtain temperatures of $123(20)\,\mu\text{K}$ and $90(20)\,\mu \text{K}$ (Fig.~\ref{fig_4}(a)). The similar temperatures rule out heating-induced loss. For comparison, $X-B$ light heats the molecules to $310(30)\,\mu\text{K}$ in 10\,ms in the absence of $\Lambda$-cooling. 

We next investigate loss into undetected rotational states, which can be separated into parity-conserving and parity-changing loss. One mechanism of parity-conserving loss arises even under $\Lambda$-cooling alone. Molecules off-resonantly excited to $A^2\Pi_{1/2}\,(v=0,J=3/2,5/2,+)$ can decay into $X^2\Sigma\,(v=0,N=3)$ (Fig.~\ref{fig_4}(b)). We find that adding a rotational repumping laser addressing the $X^2\Sigma\,(v=0,N=3)\rightarrow B^2\Sigma\,(v=0,N=2)$ transition greatly reduces this rotational loss to negligible levels~\cite{supplement}. We verify that removing this rotational repumping laser does not significantly increase the excess loss during bichromatic imaging~\cite{supplement}. We note that vibrational-changing loss into $X^2\Sigma\,(v=1,N=3)$ should be negligible because of the highly diagonal Franck-Condon factors of the $A^2\Pi_{1/2}\,(v=0) \rightarrow X^2\Sigma(v=0)$ and $B^2\Sigma\,(v=0) \rightarrow X^2\Sigma(v=0)$ transitions~\cite{supplement}.

We next quantify parity-changing rotational loss, which could arise from stray electric fields mixing in excited states of opposite parity or molecules decaying via the two-photon pathway $B\rightarrow A \rightarrow X$ (Fig.~\ref{fig_4}(d)). The admixture of opposite-parity states due to stray electric fields is expected to be much higher for the $A ^2\Pi_{1/2} (v=0,J=1/2,+)$ state compared to that for the $B ^2\Sigma (v=0,N=0)$ state because of the much smaller $A ^2\Pi_{1/2}$ $\Lambda$-doublet splitting compared to the $B ^2\Sigma$ rotational splitting separating opposite parity states. In combination with the much higher scattering rate of $X-A$ cooling photons, we conclude that the excess loss during bichromatic imaging cannot be due to stray electric fields.

To measure the loss due to two-photon decay, which results in molecules accumulating in the opposite parity rotational states $X ^2\Sigma (v=0,N=0,2)$, we directly track the $X^2\Sigma\,(v=0, N=0,F=1)$ population during bichromatic imaging using a sequence of resonant blow-out pulses, microwave sweeps and optical pumping pulses~\cite{supplement}.

As a function of the imaging duration $t$, the $X^2\Sigma\,(v=0, N=0,F=1)$ population initially rises before decaying slowly (Fig.~\ref{fig_4}(c)). Using a rate-equation model with parameters measured separately~\cite{supplement}, we extract a $B\rightarrow A$ branching ratio of $7.9(16)\times10^{-5}$, slightly lower than the theoretically predicted value of $1.3\times10^{-4}$~\cite{Hao2019dipole}, but much higher than previously measured in free-space in a molecular beam ($9(2)\times10^{-6}$)~
\cite{Truppe2017chirp}. The branching ratio indicates an $A-B$ transition dipole moment of $|d_{AB}|=0.30(3) e a_0$, which is $0.07\,e a_0$ smaller than theoretically predicted~\cite{Hao2019dipole}.

Lastly, we investigate loss that arises from admixtures of the the higher-lying $C$ states into the excited $B$ states due to the optical tweezer light. Although the trapping light is far off-resonant from any optical transition involving the ground $X^2\Sigma(v=0)$ state, it is relatively closed-detuned to the $B-C$ transitions. Sufficient coupling between the $B$ and $C$ states can give rise to observable losses through two mechanisms. First, strong anti-trapping of molecules in the $B$ state can lead to heating during bichromatic imaging and subsequent loss of molecules out of the tweezer traps. This anti-trapping occurs since the tweezer light is blue-detuned from the strongest $B^2\Sigma(v=0)-C^2\Pi(v=0,1,2)$ transitions. We have, however, already ruled out heating from our thermometry measurements. Second, admixture of the $C$ state into the $B$ state can lead to decay into unaddressed rotational and vibrational states during imaging, since the $C$ state can decay via multi-photon pathways and the $X-C$ Franck-Condon factors are non-diagonal (Fig.~\ref{fig_4}(g)).

At first glance, the theoretically predicted value of the $B-C$ transition dipole moment $|d^{th}_{BC}|=7.33\,e a_0$~\cite{Raouafi2001CaFdipole} suggests that at our tweezer wavelength, the polarizabilities of the $X$ and $B$ state differ in magnitude by $\sim 20$, and the $B$ state experiences a large and opposite ac Stark shift. Specifically, at the tweezer depth used for imaging, the predicted peak differential ac Stark shift is $\Delta^{th}_{XB} \approx h \times 600\,\text{MHz}$. The corresponding imaging loss rate into dark rovibrational states is predicted to be 6 times higher than what we observe. This motivates a careful measurement of $|d_{BC}|$. We do so by measuring the differential ac Stark shift between the $X$ and $B$ states $\Delta_{XB}$, which is sensitive to $d_{BC}^2$. $\Delta_{XB}$ is measured via loss spectroscopy by illuminating the molecules with light on the $X-B$ transition. Peak loss is observed when the $X-B$ light is on resonant with the Stark-shifted transitions. To provide optical cycling during loss spectroscopy, hyperfine sidebands addressing the four $X^2\Sigma(v=0,N=1)$ levels are added to the $X- B$ light~\cite{Holland2021GSA}. With tweezer light at 781.1\,nm, the frequency that gives maximum loss varies linearly with tweezer intensity, as shown in Fig.~\ref{fig_4}(e). Taking into account that the observed differential ac Stark shift is averaged over the trap, we infer a peak differential ac Stark shift of $\Delta_{XB} = h \times 78(2)$\,MHz at the center of the tweezer traps. This is much smaller than $\Delta^{th}_{XB}$ and provides a first indication that $|d_{BC}|$ is smaller than predicted. When the tweezer light is adjusted to 782.0\,nm, we observe minimal dependence of $\Delta_{XB}$ on the tweezer intensity (Fig.~\ref{fig_4}(e)). 

We next keep the tweezer intensity fixed at the imaging intensity and vary the wavelength of the tweezer light around 781\,nm. As shown in Fig.~\ref{fig_4}(f), the differential ac Stark shift $\Delta_{XB}$ displays two dispersive features, which originate from the $B^2\Sigma (v=0) - C^2\Pi_{1/2} (v=3)$ and $B^2\Sigma (v=0) - C^2\Pi_{3/2} (v=3)$ transitions. The amplitudes of the dispersive features directly measure $f_{\text{BC,03}} d_{BC}^2$, where $f_{\text{BC,03}}$ is the Frank-Condon factor between $B^2\Sigma(v=0)$ and $C^2\Pi(v=3)$ states. Fitting the dispersive lineshapes and using $f_{\text{BC,03}}$ calculated from spectroscopic data~\cite{Kaledin1999XABCaF,Gittins1993CaFCD}, we find a $B-C$ transition dipole moment of $|d_{BC}|=2.0(1)\, e a_0$, much smaller than the predicted value of $|d^{th}_{BC}|=7.33\,e a_0$~\cite{Raouafi2001CaFdipole}. 

Assuming that $C$-state decays always lead to loss, the admixture of the $C$ state into the $B$ state leads to an additional loss rate of $\beta^2 \Gamma_C P_B$, where $\beta$ is the amplitude of the $C$-state admixture, $P_B$ is the fraction of molecules excited to the $B$ state, and $\Gamma_C$ is the decay rate of the $C$ state. $\beta$ can be determined from the measured ac Stark shifts, and $P_B = \Gamma_{sc}/\Gamma_B$, where $\Gamma_{sc}$ is the measured $X-B$ scattering rate and $\Gamma_B$ is the $B$-state linewidth. We estimate $\Gamma_C$ by assuming that $C^2\Pi$ molecules decay only via E1 transitions, i.e. $\Gamma_{C} = \Gamma_{C,\text{rad}}$~\cite{supplement}. Using the measured value of $|d_{BC}|$, along with theoretical values of $|d_{XC}|$ and $|d_{AC}|$, we determine a radiative decay rate $\Gamma_{C,\text{rad}}$ of $2\pi \times 5.3(2)\,\text{MHz}$. The resulting estimated excess loss rates are much smaller than observed, as shown in Fig.~\ref{fig_4}(h). 

Away from the two $B^2\Sigma(v=0)- C^2\Pi(v=3)$ resonances, we only account for $\sim 30\%$ of the observed excess loss rate $\gamma_{\text{excess}}$. This suggests that either $f_{BC,03}$ is different than calculated or that additional loss mechanisms exist. The former is unlikely since extensive spectroscopy has been performed on the $X-A$, $X-B$ and $X-C-D$ systems of CaF~\cite{Kaledin1999XABCaF,Gittins1993CaFCD}. In addition, away from the two dispersive features, the background value of $\Delta_{XB}$ is only weakly sensitive to the $B-C$ Franck-Condon factors, and is consistent with our extracted value of $d_{BC}$. Possible additional loss mechanisms include pre-dissociation of the $C$ states, which would lead to $\Gamma_C>\Gamma_{\text{C,rad}}$, and photoionization of $B$-state molecules that could arise for example from absorption of an $X-A$ photon and a tweezer photon~\cite{Rice1985CaFIonization}. $B$-state molecules could also be off-resonantly excited into high-lying Rydberg states that have large decay rates~\cite{Harris1993RydbergCaF}, potentially giving rise to appreciable imaging losses.

In summary, we have demonstrated a background-free bichromatic imaging scheme for single CaF molecules in an array of $\mu$m-sized optical tweezer traps. We achieve an imaging fidelity and non-destructive detection fidelity sufficient for small-scale rearrangeable molecular tweezer array experiments. Our scheme could be extended to other laser-coolable molecules, and could be useful in experiments where stray laser-cooling light is difficult to avoid. To understand the limitations of bichromatic imaging, we have characterized multiple loss mechanisms. We note that the explored mechanisms are not specific to bichromatic imaging, and are generically relevant to fluorescent detection of trapped molecules. Lastly, our investigations have provided a new measurement of the $A-B$ transition dipole moment and the first measurement of the $B-C$ transition dipole moment of CaF. The latter was obtained through a novel method relying on dispersive ac Stark shifts, which allows measurement of transition dipole moments between excited states with arbitrarily small Frank-Condon factors. Extensions of this method could potentially be applied to investigate optical cycling in polyatomic molecules~\cite{Mitra2022arene,Zhu2022aromatic}. 

\begin{acknowledgments}
We thank John Doyle and his group for fruitful discussions and a careful reading of the manuscript. This work is supported by the National Science Foundation under Grant No. 2207518. 
\end{acknowledgments}

\bibliographystyle{apsrev4-1}

\end{document}


\title{Supplemental Material for ``Bichromatic Imaging of Single Molecules in an Optical Tweezer Array" }
\affiliation{Department of Physics, Princeton University, Princeton, New Jersey 08544, USA}
\author{Connor M. Holland}
\affiliation{Department of Physics, Princeton University, Princeton, New Jersey 08544, USA}
\author{Yukai Lu}
\affiliation{Department of Physics, Princeton University, Princeton, New Jersey 08544, USA}
\affiliation{Department of Electrical and Computer Engineering, Princeton University, Princeton, New Jersey 08544, USA}
\author{Lawrence W. Cheuk}
\affiliation{Department of Physics, Princeton University, Princeton, New Jersey 08544, USA}

\date{\today}
\maketitle

\section{Apparatus Description}
\subsection{Overview}
The experimental apparatus consists of three vacuum chambers connected in series, which we call the molecular source chamber, the intermediate chamber, and the science chamber (details in \cite{Lu2021Ring}). A cryogenic buffer gas beam (CBGB) of CaF is first generated in the molecular source chamber and passes through the intermediate chamber into the science chamber. The molecular beam is optically slowed and captured into a magneto-optical trap (MOT) at the center of the science chamber. Following magneto-optical trapping, molecules are transferred into a large-volume attractive optical trap, and ultimately into the optical tweezers, all with the aid of $\Lambda$-cooling. We describe these steps in detail in the following sections.  

\subsection{Description of Laser Beams Used}
To aid the discussion in subsequent sections, we describe the laser beams used for slowing, magneto-optical trapping, sub-Doppler cooling and fluorescent imaging.  The slowing beam consists of light  addressing the $X\, ^2\Sigma (v=0, N=1) \rightarrow B\,^2\Sigma (v=0,N=0)$ transition ($\lambda=531.0\,\text{nm}$). Sidebands addressing the ground state hyperfine manifolds are created via an electro-optical modulator (EOM) driven by a waveform created via computer-generated holography ~\cite{holland2021gsa}. The slowing beam also contains the $v=1,2,3$ vibrational repumping light ($X\, ^2\Sigma(v=1,2,3,N=1) \rightarrow A\, ^2\Pi_{1/2}(v=0,1,2, J=1/2,+)$). For the vibrational repumpers, hyperfine sidebands are generated using overdriven electro-optical modulators (EOMs).

\begin{figure}[h!]
	\includegraphics[width=1.0 \columnwidth]{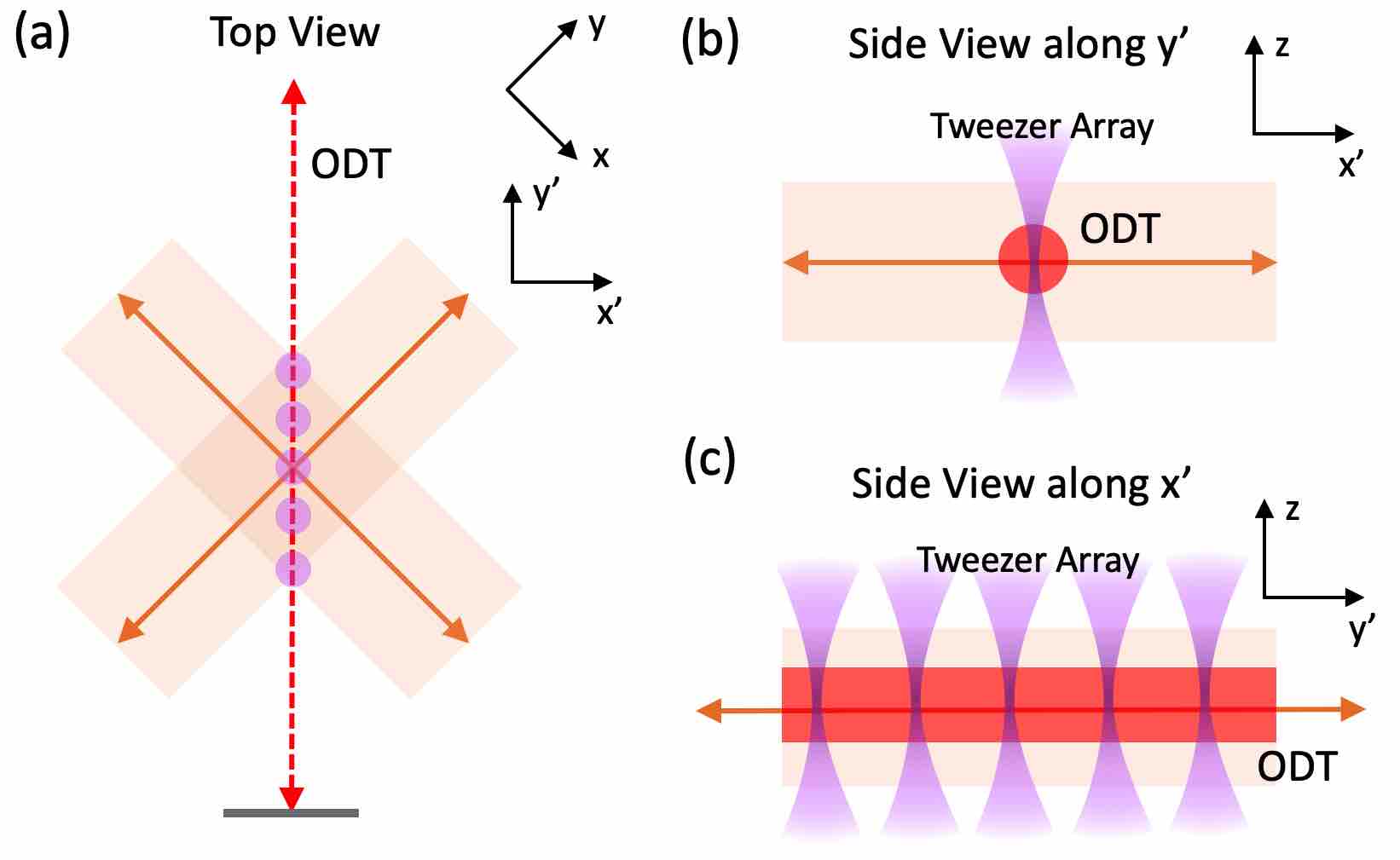}
	\vspace{-0.30in}
	\caption{\label{fig:geometry} Geometry of Various Laser Beams for Cooling and Trapping.}
	  
\end{figure}

The main cooling light ($X\, ^2\Sigma(v=0,N=1) \rightarrow A\, ^2\Pi_{1/2}(v=0, J=1/2,+)$) is used for magneto-optical trapping ~\cite{Anderegg2017MOT,Truppe2017Mot} and $\Lambda$-cooling/imaging~\cite{Cheuk2018Lambda}. This is sent via six laser beams propagating along $\pm \hat{x}$,\,$\pm \hat{y}$,\,$\pm \hat{z}$. Sidebands addressing the hyperfine manifolds are added using acousto-optical modulators (AOMs). These six laser beams  also contain $v=1$ vibrational repumping light ($X\, ^2\Sigma(v=1,N=1) \rightarrow A\, ^2\Pi_{1/2}(v=0, J=1/2,+)$) with hyperfine sidebands added using EOMs. The cooling beam propagating along $-\hat{z}$ passes through a 70:30 beamsplitter and exits the high-resolution microscope objective as a collimated beam. During magneto-optical trapping, $\Lambda$-cooling and imaging, $v=2,3$ vibrational repumping light ($X\, ^2\Sigma^+(v=2,3,N=1) \rightarrow A\, ^2\Pi_{1/2}(v=1,2, J=1/2,+)$) enter along the molecular beam axis $\hat{x}'$, which is oriented 45 degrees with respect to $\hat{x}$ and $\hat{y}$ (Fig.~\ref{fig:geometry}). 

\section{Preparing Single Molecules in Optical Tweezer Traps}

\subsection{Magneto-Optical Trapping and Sub-Doppler cooling}
We start with a molecular beam of CaF molecules generated in a cryogenic buffer gas source (CBGB)~\cite{Hutzler2012CBGB} and it is slowed on the $X-B$ transition via chirp slowing~\cite{Truppe2017chirp}. Slowed molecules are then loaded into a DC magneto-optical trap (MOT)~\cite{Truppe2017Mot}, which is subsequently compressed~\cite{Williams2018MagTrap,Ding2020YOSD}. The molecules are then released and laser-cooled to sub-Doppler temperatures via $\Lambda$-cooling~\cite{Cheuk2018Lambda,Ding2020YOSD,Langin2021SrFODT}. Details can be found in the supplemental material of~\cite{Lu2021Ring}.

\subsection{ODT Loading and Transport} 
After the molecules are magneto-optically trapped, we switch off the MOT and simultaneously turn on an attractive optical dipole trap (ODT) formed with 1064\,nm light. The beam is focused to a Gaussian waist of $w_0= 60(7)\,\mu\text{m}$, providing a single-pass trap depth of $V=k_B \times 170(30)\,\mu\text{K}$. To increase the total trap depth, the beam is retro-reflected to form a 1D lattice using a cat-eye setup. Following release from the MOT, we apply $2$\,ms of $\Lambda$-cooling at high intensities and small blue detunings. Subsequently, the $\Lambda$-cooling parameters are then adjusted to those optimal for loading molecules into the ODT. Over the next $\sim 100\,\text{ms}$, molecules are loaded into the ODT in the presence of $\Lambda$-cooling~\cite{Anderegg2018ODT,Cheuk2018Lambda,Wu2021YOODT,Langin2021SrFODT}.

Maximal transfer of molecules from the $\Lambda$-cooled cloud to the ODT is achieved when the ODT overlaps with the center of the $\Lambda$-cooled cloud. 
The location of the optical tweezer array, on the other hand, is defined by the optical axis and focal plane of the microscope objective, which is displaced from the MOT center by $\approx 500~\mu \text{m}$ along $\hat{x}'$ and $\hat{z}$ (Fig.~\ref{fig:geometry}). Since this distance is much larger than the radial width of the ODT-trapped molecules ($1/e^2$ width of $\sim 20 \,\mu\text{m}$), a transport step is needed.  

To transport the molecules, we directly move the position of the ODT after it has been loaded with molecules. In detail, the ODT beam is sent through a 2-axis acousto-optical deflector (AOD) before entering the chamber. By ramping the frequencies of the rf tones driving each axis of the AOD, we move the ODT from its initial location, centered on the MOT, to its final location in the focal plane of the optical tweezer array. The rf frequencies are chirped linearly over 30\,ms.

\subsection{Loading Molecules into Optical Tweezer Traps}
To load molecules into the optical tweezer traps, the optical tweezers are turned on following ODT transport at a depth of $V=k_B \times 0.68(12)\,\text{mK}$. $\Lambda$-cooling light is then applied using the MOT beams along $\hat{x}$, $\hat{y}$, $\hat{z}$ for  75\,ms. Subsequently, the ODT power is ramped down over 5\,ms and molecules that are not loaded into tweezers fall away over the next 15\,ms. The optimal parameters for the $\Lambda$-cooling light during tweezer loading is separately optimized. The single-photon detuning is $\Delta=39\,\text{MHz}$, and the two-photon detuning is $\delta=0\,\text{MHz}$. The hyperfine power ratio between the $F=2$ and $F=1^-$ hyperfine components is $R_{2,1}=2.1$.

\section{Generating an Array of Optical Tweezer Traps}
The optical tweezer traps are generated from light near 780\,nm, focused to near-diffraction-limited spots through a microscope objective with maximum numerical aperture of 0.65. For all data presented, we use a 1D array of 20 optical tweezers, formed from 20 regularly-spaced diffracted spots from an AOD. The frequency spacing for the AOD is set at 1\,MHz, which corresponds to an optical tweezer spacing of 4.1\,$\mu$m. 

The powers of the optical tweezer beams are balanced for uniformity. We measure the intensities of the tweezer beams at an intermediate image plane before they enter the objective, and adjust the intensities by changing the power of the RF tones  fed to the AOD. The measured intensities at the intermediate plane are balanced to better than  $1\%$. The balancing is separately verified by differential ac Stark shift measurements on the $X-B$ transition, which show a standard deviation of $\approx 10\%$ over all 20 tweezers. This is limited by the precision of the differential ac Stark shift measurements.

\section{Optical Tweezer Trap Characterization}
The optical trap depth and optical beam waist are determined via trap frequency measurements combined with measured beam intensities. First, we perform trap frequency measurements via parametric modulation of the trap depth, with the tweezer light wavelength set to 781.12\, nm. In detail, the molecules are first loaded at maximum tweezer trap depth. Subsequently the tweezer depths are modulated, and the depths are then lowered to allow parametrically heated molecules to exit the traps. Finally, the trap depths are ramped back to their initial value, and the occupations are detected using a 30\,ms bichromatic imaging pulse.

\begin{figure}[h]
	\includegraphics[width=1.0 \columnwidth]{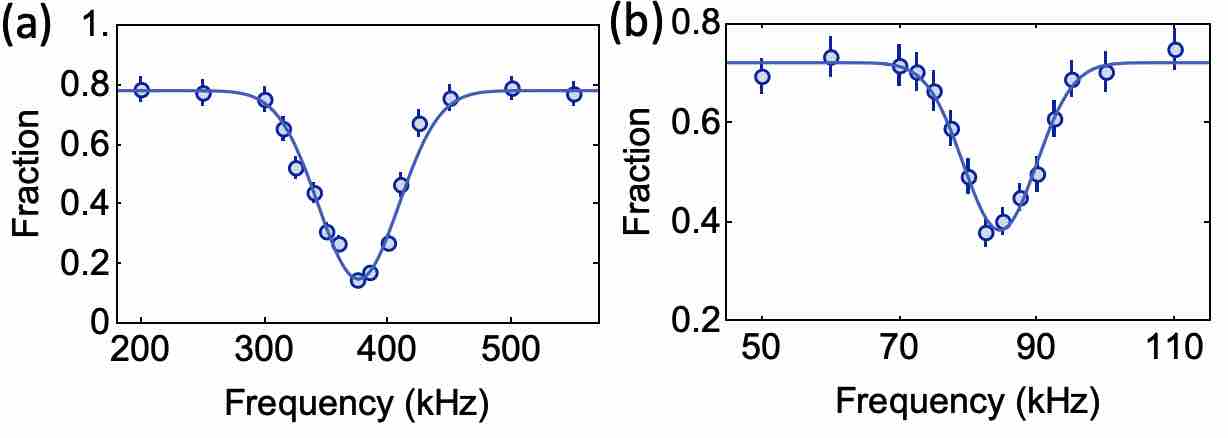}
	\caption{\label{fig:trapmod} Remaining fraction versus modulation frequency. Shown in blue are the remaining fractions following trap intensity modulation. The solid lines indicate Gaussian fits. (a) Radial frequency measurement. Peak loss is observed near twice the radial frequency. The extracted radial frequency is $\omega_r/(2\pi)$ of $187.7(6)\,\text{kHz}$. (b) Axial frequency measurement. Peak loss is observed near twice the axial frequency. The extracted axial frequency is $\omega_z/(2\pi)$ of $42.2(3)\,\text{kHz}$. }
\end{figure}

The fraction of molecules detected following trap modulation displays two loss features, which we attribute to twice the radial frequency $\omega_r$ and twice the axial frequency $\omega_z$. The measurements yield $\omega_r = 2\pi \times$187.7(6)$\, \text{kHz}$ and $\omega_z = 2\pi \times$42.2(3)$\, \text{kHz}$, where the error bars are from the fits that do not take into account systematic effects. Assuming a Gaussian beam in the paraxial approximation, the trap frequency ratio provides an estimate of the beam waist $w_0$:
\begin{equation}
w_0 = \frac{\lambda}{\pi \sqrt{2}} \frac{\omega_r}{\omega_z}. 
\end{equation}
This provides an estimate of $w_0 =$ 781(3)\,nm. However, optical aberrations could significantly affect the trap frequency ratios, and hence this estimate. While the radial profile of a beam is typically well described by a Gaussian, aberrations can significantly distort how the trap depth varies along the optical axis. Typically, these aberrations lower $\omega_z/\omega_r$ as rays at different incoming angles focus to different axial positions.

To provide a better estimate of the radial beam waist $w_0$, which is important for determining the overall ac Stark shifts and molecular temperatures, we use a second method that relies on power measurements and a calculation of the ground state polarizability of CaF. In detail, the diffracted power per tweezer beam is measured in an intermediate imaging plane following the AOD, and the transmission of the subsequent optics are taken into account. We then calculate the polarizability of the $X^2\Sigma (v=0)$ state (taking into account electronically excited A,B,C and D states) to determine the radial trap frequencies for a variety of Gaussian beam waists. The polarizability of the $X^2\Sigma (v=0)$ state is well-known, since the dominant trapping contributions are from the $A$ and $B$ states, which have been well characterized. States we exclude ($E$ and higher) are estimated to contribute $<1\%$ to the polarizability. We assume that the molecules, which are distributed among the various $N=1$ hyperfine states, on average experience the scalar polarizability. Using this method, we find a Gaussian waist of $w_0 = 720(14)\,\text{nm}$ and a trap depth of $V=k_B \times 1.28(11)\,\text{mK}$. 

The uncertainties on the preceding quantities are estimated as follows. The uncertainty in the optical power per tweezer is estimated to be $\sim 8\%$, while the uncertainty in the polarizability is estimated to be $\sim 4\%$, arising predominantly from the uncertainty in the excited state lifetime of the $A^2\Pi_{1/2}$ state~\cite{Wall2008XA}. Note that when extracting the beam waist from the radial trap frequency, the beam waist depends on the total power and polarizability as $(P\alpha)^{1/4}$, but is inversely proportional to the trap frequency. Therefore, the radial trap frequency is highly sensitive to the beam waist, but less so to the total power or the ac polarizability.

Since we have high confidence in the measured power and the calculated polarizability, and since we are primarily interested in the trap depth and radial size, throughout our work, we model the trap as an ideal Gaussian beam with beam waist $w_0 = 720(14)\,\text{nm}$ and tweezer trap depth $V=k_B \times 1.28(11)\,\text{mK}$.

\section{Detecting and Identifying Single Molecules}
\subsection{ Imaging Setup}

In the main text, two methods of imaging are described: 1) single-color $\Lambda$-imaging with $\Lambda$-cooling light on the $X-A$ transition, and 2) bichromatic imaging consisting of $\Lambda$-cooling light along with light addressing the $X\, ^2\Sigma^+ (v=0, N=1,F=0) - B\,^2\Sigma^+ (v=0,N=0)$ transition. 

For $\Lambda$-imaging, we use the six beams along $\pm \hat{x}$, $\pm \hat{y}$ and $\pm \hat{z}$. Each beam has two hyperfine components nominally addressing the $X(v=0,N=1,F=1^-,2)$ hyperfine manifolds. For bichromatic imaging, we additionally add four small beams ($\approx 2\,\text{mm}$ diameter) propagating mostly along $\pm \hat{x}$, $\pm \hat{y}$ addressing the $X\, ^2\Sigma^+ (v=0, N=1,F=0) - B\,^2\Sigma^+ (v=0,N=0)$ transition. Due to optical access constraints, these beams have a small vertical component. The plane spanned by these beams intersect the $x$-$y$ plane at $\sim 5^\circ$. 

For imaging, $X-B$ fluorescence at 531\,nm is collected through the microscope objective and imaged onto an EMCCD camera.
The microscope objective is apochromatic and has a numerical aperture (NA) of 0.65. Due to apertures behind the objective, the working NA is slightly reduced. This reduction is not axially symmetric. Using the exact geometry of the apertures, we numerically compute the reduction factor to the overall collection efficiency. To filter out the large amount of $\Lambda$-cooling light (606\,nm) and $v=1$ light (628\,nm) that propagates and scatters along $\pm \hat{z}$, two dichroic filters that pass 531\,nm light but filter out 606\,nm and 628\,nm light are placed in the imaging path. 

Taking into account the transmission of the objective, the effective aperture size, the transmissions of the vacuum window, various compensation optics, beamsplitters and dichroic beamsplitters, the overall efficiency of the photon detection is 0.049(2). This also takes into account the quantum efficiency of the camera and the reflection of the camera window.

\subsection{Double-Imaging Procedure}
\label{sec:doubleimg}
\begin{figure}[h!]
	\includegraphics[width=1.0 \columnwidth]{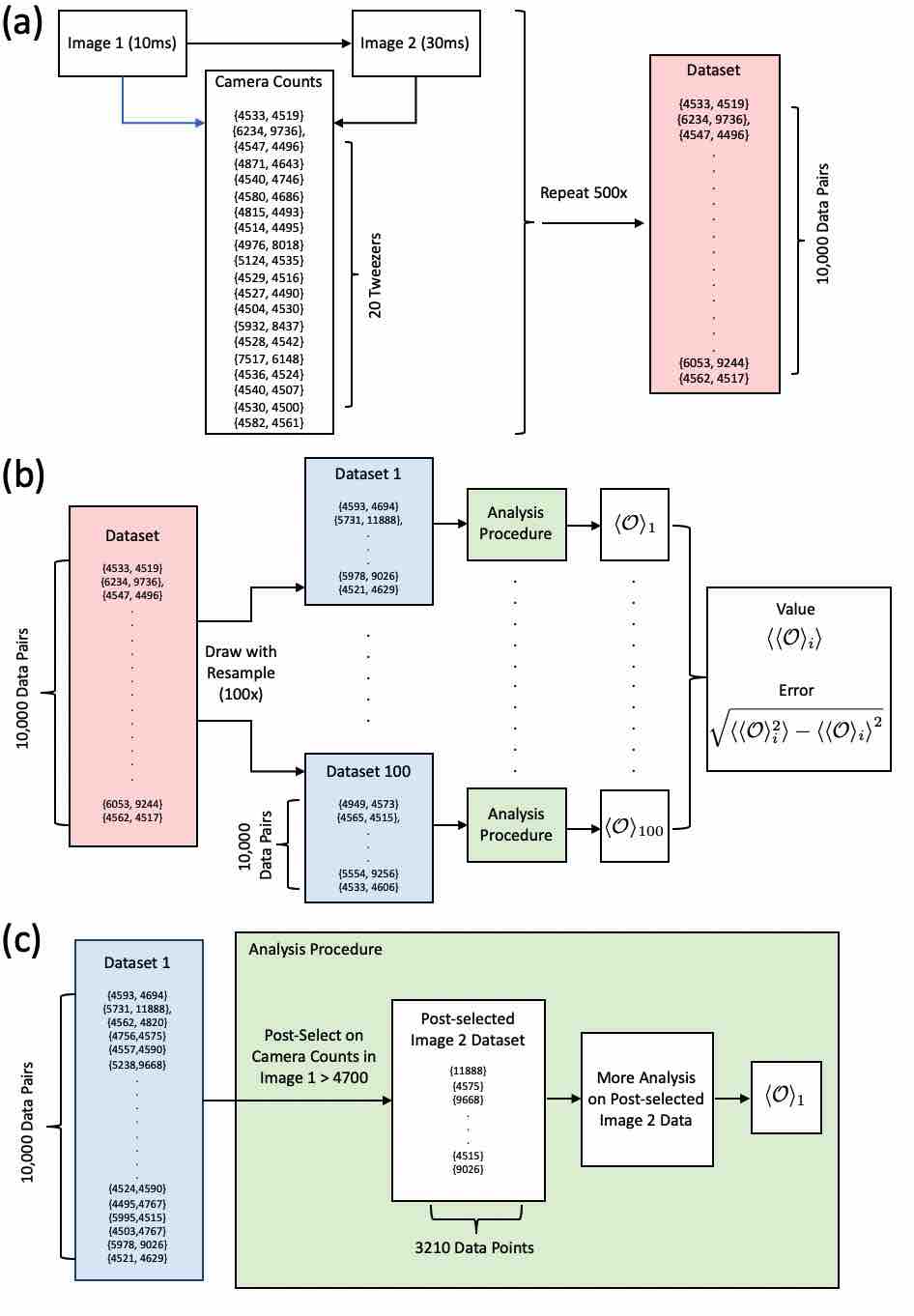}
	\vspace{-0.30in}
	\caption{\label{fig:analysisproc} (a) Double Imaging Procedure. For each experimental run, the molecules are imaged twice. The exposures for the first and second image are 10\,ms and 30\,ms, respectively. By summing the total camera counts in each 3x3 pixel region around each tweezer, each experimental run produces 20 pairs of numbers. We treat each tweezer as equivalent. By repeating the experiment 500 times, we obtain a dataset of 10,000 pairs of total camera counts.
	(b) Bootstrapping Procedure. We use bootstrapping to estimate errors. From the initial dataset with 10,000 pairs of numbers, we create 100 new datasets by drawing pairs randomly from the initial dataset, with resampling. Each boot-strapped dataset also has 10,000 pairs. Any analysis procedure is applied to all boot-strapped datasets. For example, for an analysis procedure that produces an observable $\bra \mathcal{O} \ket$ for a dataset, we obtain 100 values $\bra \mathcal{O}\ket_i$ for the 100 boot-strapped datasets. The final reported value of the observable is the mean of $\bra \mathcal{O}\ket_i$, and the reported error is their standard deviation.
	(c) Example Post-Selection Procedure. For each bootstrapped dataset, we post-select on the total camera counts from image 1, which creates a smaller dataset for further analysis.   
	  }
\end{figure}

For each experimental run, we image the molecules twice sequentially using bichromatic imaging (Fig.~\ref{fig:analysisproc}(a)). The first image has an exposure time of 10\,ms, and is intended to be mostly non-destructive. The second image has an exposure time of 30\,ms, and is intended to provide optimal detection fidelity. As described in Section~\ref{sec:fidelity}, post-selection on the first image allows us to estimate classification errors. This double-imaging procedure also allows normalization of the average molecule number, which is crucial for the population dynamics measurements and the ac Stark shift measurements. The post-selection process removes the effect of number fluctuations and photon shot noise. The only remaining noise is molecular number shot noise.

\subsection{Imaging Histograms}
To classify whether an optical tweezer trap is occupied, for each second image, we crop a square region ($3\times3$ binned pixels, binning 2x2, 2.2\,$\mu$m x 2.2\,$\mu$m) about each tweezer location (20 tweezers per image). Next, we sum the total camera counts within each region. The 20 tweezers are assumed to be equivalent. From multiple images (total of 500), we can create histograms of camera counts for both the first image (10\,ms exposure) and second image (30\,ms exposure). The histogram created from the second image (30\,ms exposure) is shown in Fig.~1(d) in the main text. For both images, the resulting histograms are bimodal, with the lower peaks corresponding to unoccupied tweezers and the higher peaks corresponding to occupied tweezers. One can then set a threshold $\theta$ above which a tweezer is classified to be occupied. Although the histograms are bimodal, the overlap between the two peaks leads to classification errors, which we estimate using an analysis procedure described in Section~\ref{sec:fidelity}. Note that in the main text, only the histogram for the second image (30\,ms exposure) is used to obtain the imaging fidelity (30\,ms exposure).  Also note that in the main text, the non-destructive detection fidelity is directly measured, and we do not need to create histograms or obtain classification errors for the first image (10\,ms).

\section{Fidelity Estimation}
\label{sec:fidelity}
In this section, we describe in detail how we extract classification errors, the imaging fidelity and the non-destructive detection fidelity. 

\subsection{Classification Errors}
Classification errors are of two types: occupied tweezer sites identified as empty, and empty sites identified as occupied. We denote these errors as follows: $\epsilon_{10}$ is the probability of classifying an occupied tweezer site (at the beginning of the imaging pulse) to be empty; $\epsilon_{01}$ is the probability of classifying an empty tweezer site (at the beginning of the imaging pulse) to be occupied. We note that these errors depend on the threshold $\theta$ used for classification. 

From the two errors, $\epsilon_{10}$ and $\epsilon_{01}$, one finds that the total error $\epsilon$ is given by $\epsilon = p \epsilon_{10}+(1-p)\epsilon_{01}$. This is dependent both on the the threshold $\theta$ and the true tweezer occupation probability $p$.

\subsection{Definition of Imaging Fidelity}
We first define the occupation-dependent detection fidelity $f_\text{det}(p)=\text{Max}_{\theta}( 1-\epsilon)$ as the maximum fidelity (minimum total error) achieved for a given a tweezer occupation probability $p$, optimized over all thresholds $\theta$. We note that both the detection fidelity $f_\text{det}(p)$ and the optimal threshold is also a function of $p$. For example, for $p=1$, $f_\text{det}(1)=1$ can be achieved with $\theta=0$. To provide an occupation-independent metric for detection fidelity, we define the imaging fidelity $f$ as the detection fidelity obtained for $p=0.5$, i.e. $f=f_\text{det}(p=0.5)$. We note that $f_\text{det}(0.5)$ is typically lower than $f_\text{det}(p\approx 1)$ or $f_\text{det}(p\approx 0)$. 

\subsection{Definition of Non-Destructive Detection Fidelity}
A second metric of imaging performance relevant to rearrangeable tweezer arrays is the non-destructive detection fidelity  $f_\text{ND}$, which we define as the probability that a trap classified to be occupied is truly occupied. The error $\epsilon_{ND}=1-f_\text{ND}$ is also known as the retention error~\cite{Graham20192Dgate}. This error can be written as
\begin{equation}
\epsilon_{ND}=1-\frac{p(1-\epsilon_{10}) f_{\text{surv}}}{p(1-\epsilon_{10} )+  (1-p)\epsilon_{01}}
\label{epsilonND}
\end{equation}
where $f_{\text{surv}}$ is the survival probability following non-destructive detection. For high detection fidelities and survival probabilities, 
\begin{equation}
\epsilon_{ND} \approx \left(\frac{1}{p}-1\right)\epsilon_{01}
\end{equation} 
This means that the retention error can be minimized by reducing $\epsilon_{01}$, which can be achieved by using a higher threshold $\theta$. A higher threshold, however, comes at the expense that an occupied tweezer has a higher probability of being misclassified as unoccupied, which reduces the data rate for single tweezers and effectively lowers the tweezer loading rate for tweezer arrays. This effect can be captured by the data rejection rate $f_R$, defined as the probability that an occupied tweezer will be classified as unoccupied. $f_R$ increases as the threshold increases, and is given by
\begin{equation}
f_R =  \epsilon_{10}
\end{equation}

\subsection{Error Model and Imaging Fidelity Extraction}
\label{sec:fidext}
To estimate the imaging fidelity, we first determine the errors $\epsilon_{10} (\theta)$ and $\epsilon_{01}(\theta)$ as a function of the threshold $\theta$. The errors $\epsilon_{10} (\theta)$ and $\epsilon_{01}(\theta)$ can be determined if we know the histogram for an empty tweezer and the histogram for an occupied tweezer. While one can create empty tweezers by not loading molecules, creating occupied tweezers with $100\%$ probability directly is not possible, since the tweezer loading process is stochastic. 

To obtain the histogram for an occupied tweezer, we make use of the double-imaging procedure described in Section~\ref{sec:doubleimg}, followed by a subtraction procedure during image analysis. For each experimental run, we load molecules into the 20 tweezers stochastically, and then take two consecutive images (Fig.~\ref{fig:analysisproc}(a)). The two images have imaging durations of 10\,ms and 30\,ms, respectively, with the first image intended to be minimally destructive. We next create a series of post-selected histograms from the second image conditioned upon camera counts exceeding a series of classification threshold $\theta_1$ in the first image (Fig.~\ref{fig:analysisproc}). When creating histograms, we treat all 20 tweezers to be identical. Different values of $\theta_1$ lead to different post-selected histograms corresponding to different tweezer loading rates. 

\begin{figure}[h!]
	\includegraphics[width=1.0 \columnwidth]{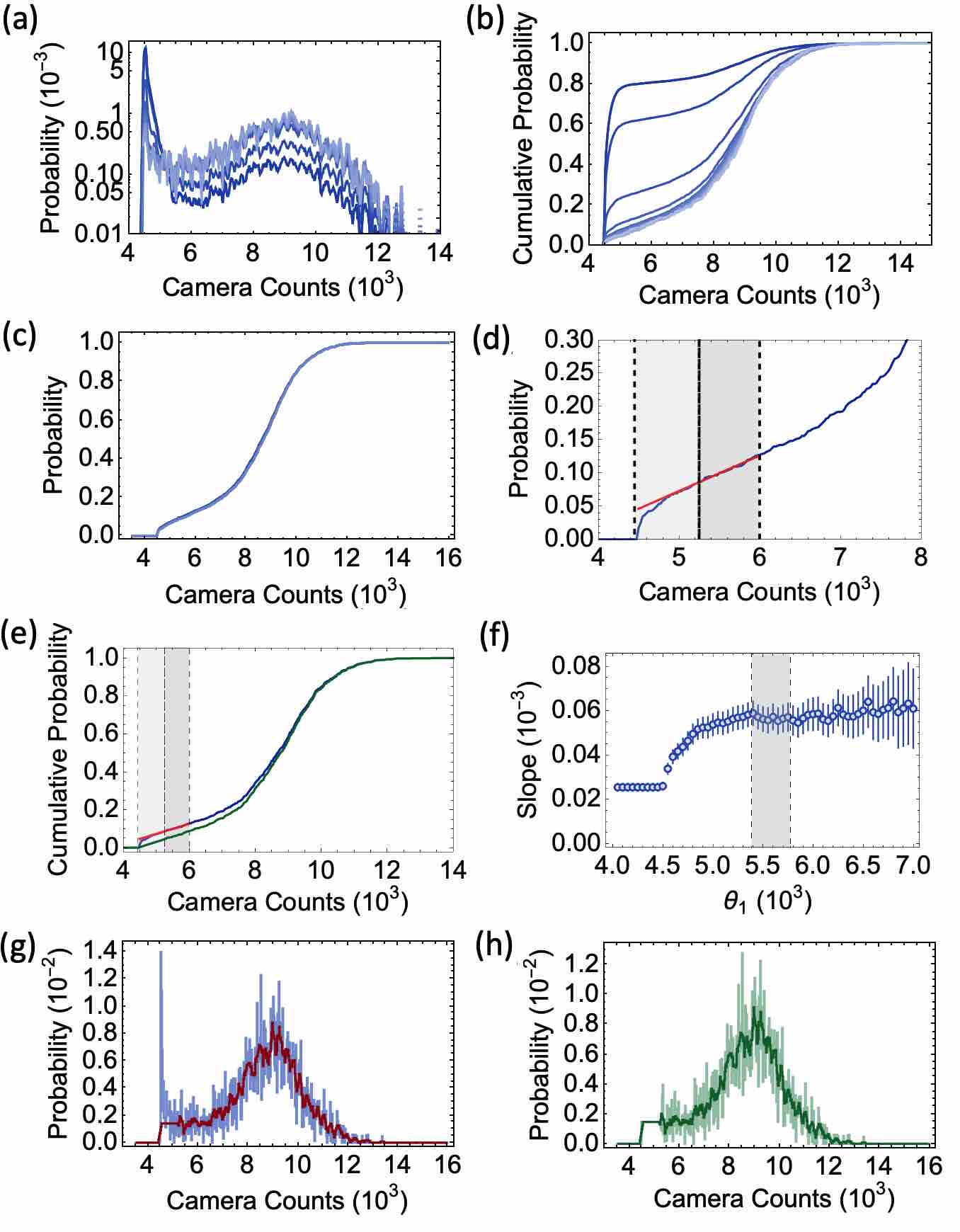}
	\vspace{-0.30in}
	\caption{\label{fig:fidelitypanel} (a) Normalized histograms from the second images versus camera counts $\theta_2$. The curves correspond to different post-selection threshold $\theta_1$ used in the first image, which controls the effective loading rate in the post-selected datasets. (b) Cumulative probability functions $g(\theta_2)$ as a function of camera counts $\theta_2$ in the second image, corresponding to the curves in (a). As $\theta_1$ increases, the histograms have less contribution from unoccupied tweezers, but the noise grows as the post-selected dataset becomes smaller. In (a,b), from dark to light, $\theta_1$ ranges from 4050 to 6800 in steps of 250. (c) Cumulative probability curves $g(\theta_2)$ for the range of post-selection parameter $\theta_1$ that is used to extract the histograms for occupied tweezers. (d,e) Subtraction procedure. In (d), data in the right shaded region is fit to a linear curve, shown in red. Data in the left shaded region is discarded and replaced by the linear fit, producing $g'(\theta_2)$. The subtracted distribution is shifted down and renormalized, producing $g''(\theta_2)$, shown in green in (e). (f) Sensitivity of linear fitting. The fitted slope in (e) is shown as a function of post-selection threshold $\theta_1$. To obtain the true distribution for occupied sites, $g''(\theta_2)$ are averaged for $\theta_1$ in the shaded gray region. (g) The post-selected histogram averaged over the range of $\theta_1$ indicated in (e) is shown in blue. The subtracted histogram, before normalization, is shown in red. A moving average of 5 bins is applied for clarity. (h) The subtracted and renormalized histogram, given by the derivative of $g''(\theta_2)$, is shown in light green. The green line shows the curve with a moving average of 5 bins.}
\end{figure}

As shown in Fig.~\ref{fig:fidelitypanel}(a), the post-selected histograms are bimodal. At sufficiently large thresholds $\theta_1$, the post-selected normalized histograms display a suppressed lower peak. However, a small peak corresponding to a small fraction of unoccupied sites is still visible. In order to obtain the histogram from an occupied tweezer, we subtract off this small contribution with the following procedure. 

Over a range of $\theta_1$ where the lower peak is suppressed but sufficient data remains to provide acceptable signal-to-noise ratios, we create normalized cumulative distribution functions (CDFs) $g(\theta_2)$ from the histograms. As shown in Fig.~\ref{fig:fidelitypanel}(b), one observes that the CDF rises abruptly at a threshold $\theta_{\text{thres}}$, and then increases approximately linearly above the threshold. This linear portion corresponds to a flat histogram distribution near zero photon counts, which arises because of the finite imaging lifetime~\cite{Cooper2018AEA}. The sharp rise near $\theta_{\text{thres}}$ in the CDF corresponds to the small unoccupied peak. We remove this contribution by fitting the linear portion of the CDF, and replacing the non-linear portion near $\theta_{\text{thres}}$ with a fitted linear curve truncated at $\theta_{\text{thres}}$ (Fig.~\ref{fig:fidelitypanel}(d)). As shown in Fig.~\ref{fig:fidelitypanel}(e), this distribution, $g'(\theta_2)$, contains a jump at $\theta_{\text{thres}}$, which corresponds to the area under the lower peak in the imaging histogram. In the final step, we subtract this jump and normalize the CDF to obtain the true CDF $g''(\theta_2)$ for occupied tweezers. Specifically, we subtract $g'(\theta_{\text{thres}})$ from $g'(\theta_2)$ and rescale the resulting curve such that $g''(\infty) =1$ (Fig.~\ref{fig:fidelitypanel}(e)). We note that $g''(\theta_2)$ in fact coincides with $\epsilon_{10}(\theta_2)$, the probability of classifying an occupied site as empty for a classification threshold $\theta_2$. The normalized histogram for an occupied tweezer, $h(\theta_2)$ can be obtained by taking the derivative of $g''(\theta_2)$ with respect to $\theta_2$ (Fig.~\ref{fig:fidelitypanel}(h)).

To obtain $\epsilon_{01}$, we use the histogram for an unoccupied tweezer. This can be accomplished simply by shifting the region about each tweezer to be be sufficiently far away from the tweezer position (10 binned camera pixels (2x2 binning), 7.2\,$\mu$m away). The resulting histograms simulate unoccupied tweezers with similar background light as the imaging regions for the tweezers. The function $\epsilon_{01}(\theta_2)$ can then be obtained by summing the number of counts above $\theta_2$, and normalizing by the total number of counts. 

\subsection{Estimating Uncertainties in Fidelity Estimation}
\label{sec:errcorr}
In order to estimate the uncertainties in the above procedures, we use a boot-strapping procedure. We use a dataset of 500 pairs of images, each with 20 tweezers, giving a total dataset corresponding to 10,000 tweezers. Each of the 10,000 realizations gives a pair of numbers corresponding to the total number of camera counts in the two consecutive images. We then generate 100 boot-strap datasets each with 10,000 realizations, by sampling the 10,000 realizations with replacement. The above analysis procedures are then carried out for each of these 100 boot-strap datasets. The uncertainties in the extracted error functions $\epsilon_{10} (\theta_2)$ and $\epsilon_{01}(\theta_2)$ are obtained from the standard deviation over the 100 boot-strap datasets. The bootstrapping procedure is summarized in Fig.~\ref{fig:analysisproc}(b).

\subsection{Validating the Error Model/Measurement Correction Procedure}
To validate the error model, we estimate the tweezer occupations $p$ of post-selected datasets using different classification thresholds $\theta_2$. If the error model is valid, the error-corrected occupations should not depend on $\theta_2$.

\begin{figure}[h]
	\includegraphics[width=1.0 \columnwidth]{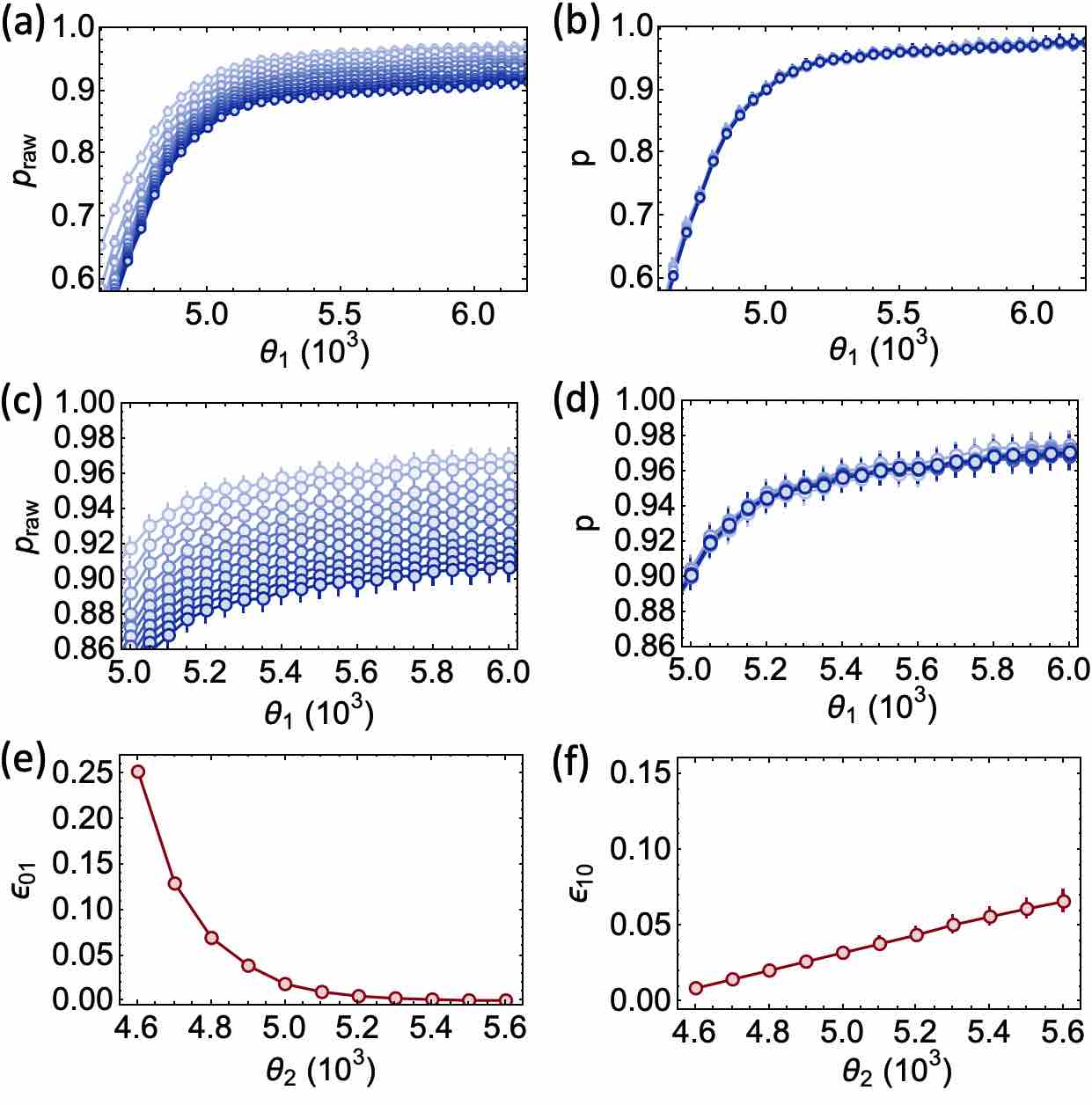}
	\caption{\label{fig:errcorr} (a) Raw occupation probability $p_{\text{raw}}$ versus post-selection criterion $\theta_1$. $\theta_1$ is the classification in the first non-destructive image, which controls the loading probability of the post-selected second images. The different curves correspond to different classification thresholds $\theta_2$ used in the second image. (b) Corrected occupation probability $p$ versus post-selection criterion $\theta_1$. All curves for different identification thresholds $\theta_2$ collapse into one curve. (c) Zoomed-in version of (a). (d) Zoomed-in version of (b). For panels (a-d), the various curves show results using different classification thresholds $\theta_2$ used on the post-selected second images. From light to dark, the classification thresholds $\theta_2 $ range from 4600 to 5600 in steps of 100. (e,f). The error functions $\epsilon_{10}(\theta_2)$ and $\epsilon_{01}(\theta_2)$ used to correct the data shown in (a,c).}
\end{figure}

In detail, we first generate post-selected datasets by conditioning upon camera counts exceeding a series of classification threshold $\theta_1$ in the first image (Fig.~\ref{fig:analysisproc}), as described previously. For each post-selected dataset, we generate a corresponding histogram in the second image. The raw occupation fraction in the second image, $p_{\text{raw}}$ is  obtained for a variety of classification thresholds $\theta_2$. In Fig.~\ref{fig:errcorr}(a,c), we plot $p_{\text{raw}}$ as a function of 
the post-selection parameter $\theta_1$, for a variety of classification thresholds $\theta_2$. Because the post-selection parameter $\theta_1$ controls the occupation probability of the post-selected dataset, Fig.~\ref{fig:errcorr}(a,c) effectively shows $p_{\text{raw}}$ versus loading rate for a variety of tweezer occupations parametrized by $\theta_1$. One observes that different classification thresholds $\theta_2$ produces different curves, with a difference of $\sim 5-10\%$ over the range of classification thresholds used. This is because the measurement errors depend on $\theta_2$, and the $p_{\text{raw}}$ curves for different $\theta_2$ are affected differently.

We next obtain $p$ by correcting using the error functions $\epsilon_{10} (\theta_2)$ and $\epsilon_{01}(\theta_2)$ determined in the procedure described in Section~\ref{sec:fidext}. Noting that \begin{equation}
p_{\text{raw}} = p (1-\epsilon_{10}) + (1-p) \epsilon_{01},
\end{equation}
one obtains
\begin{equation}
p = \frac{p_{\text{raw}}-\epsilon_{01}}{(1-\epsilon_{10}-\epsilon_{01})}.
\end{equation}
For each curve in Fig.~\ref{fig:errcorr}(a,c), which correspond to a classification threshold $\theta_2$, we correct the curves using the error functions $\epsilon_{10} (\theta_2)$ and $\epsilon_{01}(\theta_2)$ to obtain $p$. As shown in Fig.\ref{fig:errcorr}(b,d), after correcting for classification errors, all curves with different classification thresholds $\theta_2$ collapse into a single curve within the boot-strap error bars, despite $\epsilon_{01}$ and $\epsilon_{10}$ being as large as $\sim 25\%$ at the maximum and minimum thresholds $\theta_2$ (Fig.\ref{fig:fidelitypanel}(e,f)). This validates the procedure used to estimate $\epsilon_{01}(\theta_2)$ and $\epsilon_{10}(\theta_2)$, that is, the error model is accurate.

\subsection{Measuring the Dependence of Non-Destructive Fidelity on the Data Rejection Rate}
Instead of inferring $f_{\text{ND}}$ from imaging loss and error rates in the non-destructive image via Eq.~\ref{epsilonND}, we directly measure the non-destructive fidelity. We use the experimental sequence and dataset used in extracting destructive detection fidelity. For each experimental run, we load molecules into the 20 tweezers stochastically and use the double imaging procedure described in Section~\ref{sec:doubleimg}. We next create a series of post-selected histograms from the second image, conditioned upon camera counts exceeding a series of classification threshold $\theta_1$ in the first image. When creating histograms, we treat all 20 tweezers to be identical. Different values of $\theta_1$ lead to post-selected histograms corresponding to different tweezer loading rates. 

The probability $p$ versus the first image threshold $\theta_1$ obtained in Section~\ref{sec:errcorr} directly probes $f_{\text{ND}}(\theta_1)$. 
Next, for each post-selected data set conditioned on $\theta_1$, we estimate the number of tweezers that are classified as occupied $N_{\text{true,post}}(\theta_1)$ compared to the total number of truly occupied tweezers $N_{\text{true,all}}$ before the second image. The data rejection rate $f_R$ can be written as
\begin{equation}
f_R = 1 - \frac{N_{\text{true,post}}(\theta_1) }{N_{\text{true,all}}},
\end{equation}
and captures the fraction of occupied tweezers that are not used since the first non-destructive image did not survive post-selection.

To obtain $N_{\text{true,post}}$ and $N_{\text{true,all}}$ without detailed analysis of the histograms obtained from the first image, we make use of the true normalized second-image histogram for a loaded tweezer $h(\theta_2)$. As described in Section~\ref{sec:fidext}, $h(\theta_2)$ is obtained by taking a derivative of $g''(\theta_2)$ with respect to $\theta_2$. Next, we pick a threshold $\theta_{\text{load}}$ well beyond the unoccupied peak. 
We then obtain $N_{\text{true,all}}$ by taking the ratio of the integrated area of the second image histogram above $\theta_{\text{load}}$, and dividing the result by the integrated area of $h(\theta_2)$ above $\theta_{\text{load}}$.
 $N_{\text{true,post}}(\theta_1)$ is obtained similarly on post-selected histograms parametrized by the first-image threshold $\theta_1$.
 
This procedure allows us to obtain $f_R(\theta_1)$, the data rejection rate. The dependence on the non-destructive fidelity on data rejection rate can then be obtained simply by plotting $(f_{\text{ND}}(\theta_1), f_R(\theta_1))$ parametrically, as shown in Fig.~2(b) in the main text.

\section{Thermometry}
To measure the temperature of the molecules in the tweezer trap, we perform release and recapture measurements. In detail, following the first image, where the initial occupations of each tweezer trap are determined, the tweezer traps are switched off abruptly and the molecules are released. After a variable hold time, during which the molecular samples expand, the tweezer traps are then abruptly switched on. The fraction of molecules recaptured is subsequently measured. The release-and-recapture curves are normalized to the survival fraction at zero release time.

The release-and-recapture curves are then fit to classical Monte Carlo simulations with different initial molecular temperatures. The simulations assume an ideal Gaussian trap with beam waist $w_0 = 720(14)\,\text{nm}$ and trap depth $V=k_B \times 1.28(11)\,\text{mK}$. Gravity is also included in the simulation. In detail, molecules are first initialized from a thermal distribution ignoring the small effect of gravity. The temperature of this sample is obtained by directly computing the kinetic energy, and using the Virial theorem for a harmonic trap $k_B T = \frac{1}{2} m \bra v^2 \ket$, which is expected to be valid when $k_B T\ll V$. The abrupt release and recapture sequence is then simulated. Following recapture, we evolve the system for 5\,ms to ensure that untrapped molecules fall away. We then obtain the final fraction of molecules that remain in the trap  by counting molecules remaining in the region defined by $r<3w_0, |z|<3z_R$. This produces simulated release-and-recapture curves. The experimental data is then fit to the simulated data with two free parameters, the overall scale and the temperature. The uncertainties in the trap depth and beam waist are taken into account in the following way. Separate simulated data sets are produced with either $w_0$ or $V$ offset by 1$\sigma$ from its center value. The data is then fit to these two curves. The deviations in the fitted temperature are then added in quadrature to produce error bars. 

\section{Measuring Differential ac Stark Shifts between $X ^2\Sigma $ and $B^2\Sigma$ } 
In the main text, we consider the scalar Stark shift, which dominates in our light shift measurements. In detail, we measure the Stark shift of $X^2\Sigma(v=0,N=1) - B^2\Sigma (v=0, N=0)$, averaged over all ground states, through loss spectroscopy. This is accomplished by adding sidebands to the spectroscopy light that is spectrally engineered to address all four ground state hyperfine manifolds~\cite{holland2021gsa}. When the spectroscopy light is on resonance, one can cycle photons and heat the molecules out of the tweezers. We note that the hyperfine splitting in the excited state is almost unresolved and is smaller than the observed loss features. We therefore ignore the hyperfine coupling in the excited $B^2\Sigma(v=0,N=0,J=1/2)$ manifold. To check non-linearities in intensity arising from hyper-polarizability, we measure the differential ac Stark shift  between the $X^2\Sigma$ and $B^2\Sigma$ states as a function of tweezer depth. Over the entire depth range, our measurements exhibit a linear dependence on tweezer intensity (Fig.~\ref{fig:Starkshift}). Extrapolation to zero intensity also allows extraction of the free-space transition frequency.

\section{Modeling ac Stark Shifts} 
In this section, we list contributions to the ac Stark shifts in the $X$ and $B$ states.

\subsection{General Forms for $^2\Sigma$ and $^2\Pi$ States}
We are primarily concerned with scalar ac Stark shifts from $^2\Sigma \rightarrow\, ^2\Sigma$ and $^2\Sigma \rightarrow\, ^2\Pi_{\Omega=1/2,3/2}$ transitions and use $U_{\alpha\beta}$ to denote the ac Stark shift experienced by state $\alpha$ ($^2\Sigma$ state) due to state $\beta$ ($^2\Sigma$ or $^2\Pi$ state). One finds, for the case of a $\beta ^2\Sigma$ state,
\begin{equation}\label{sigmashift}
U_{\alpha\beta} = \mp3 \pi c^2 \frac{\Gamma_\beta}{2\omega_0^3 }\frac{1}{3} f_{\alpha\beta} \left(\frac{1}{\omega_0 -\omega}+ \frac{1}{\omega_0+\omega}\right) I,
\end{equation}
and, for the case of a $\beta^2\Pi$ state (ignoring spin-orbit splitting and summing over both $\Omega$ manifolds),
\begin{equation}\label{pishift}
U_{\alpha\beta} = \mp 3 \pi c^2 \frac{\Gamma_\beta}{2 \omega_0^3 }\frac{2}{3} f_{\alpha\beta} \left(\frac{1}{\omega_0 -\omega}+ \frac{1}{\omega_0+\omega}\right)I,
\end{equation}
where $\mp$ is used if the $\beta$ state is higher(lower) in energy than the $\alpha$ state, $\omega_0$ is the resonant angular frequency, $I$ the intensity, and $f_{\alpha\beta}$ the Franck-Condon factor from state $\alpha$ to state $\beta$. We have ignored spin-rotation and hyperfine splittings to produce Eqs. \ref{sigmashift} and \ref{pishift}. Note that the pre-factor $3 \pi c^2 \frac{\Gamma_\beta}{2 \omega_0^3 }$ is proportional to $d_{\alpha\beta}^2$, where $d_{\alpha\beta}$ is the transition dipole matrix element between the two electronic states.

\subsection{$X^2\Sigma$ ac Stark Shifts}
\label{sec:Xstark}
To estimate the $X^2\Sigma$ Stark shifts, we take into account coupling to the $A^2\Pi_{1/2}$, $A^2\Pi_{3/2}$, $B^2\Sigma$, $C^2\Pi_{1/2}$, $C^2\Pi_{3/2}$, $D^2\Sigma$, $E$, and $F$ electronic states. Since the trapping wavelength is red-detuned with respect to all transitions from the $X^2\Sigma(v=0)$ states, the $X^2\Sigma(v=0)$ state experiences only negative ac Stark shifts. When computing the ac Stark shifts, we resolve vibrational states for the $A^2\Pi_{1/2}$, $A^2\Pi_{3/2}$, $B^2\Sigma$, and $C$ states, but not for the $D$, $E$ and $F$ states. In addition, because of the large detunings to $C$ and higher states, the spin-orbit splitting between $C^2\Pi_{1/2}$ and $C^2\Pi_{3/2}$ is ignored. For all ac Stark shift calculations, the spin-rotation and hyperfine splittings are ignored. To calculate the ac Stark shifts due to each vibrational line, we use Franck-Condon factors calculated using spectroscopic constants from~\cite{Kaledin1999XABCaF,Gittins1993CaFCD,Hao2019dipole}. 

We note that at the trapping wavelength of 781.1\,nm, the $A^2\Pi_{1/2,3/2}$ and $B^2\Sigma$ states account for $ 96.6\%$ of the total ac Stark shift for the $X^2\Sigma$ state since they are the nearest detuned. The $C$ states account for $ 3\%$ of the total ac Stark shift, the $D$ state $ 0.1\%$, the $E$ state $= 0.05\%$, and the $F$ state $ 0.2\%$. We use experimentally measured transition dipole moments for the $X-A$ and $X-B$ transitions~\cite{Dagdigian1974XB,Wall2008XA}, and theoretically calculated transition dipole moments for all other transitions except for the $X-F$ transition~\cite{Raouafi2001CaFdipole}, which is not available in the literature. We use the transition dipole moment of $X-E$ for the $X-F$ transition, which is likely an overestimate. The $X-F$ contribution is negligible and well below the experimental uncertainties. We also ignore variations in transition dipole moments with vibrational quantum number, which is typically small. In addition, since the Franck-Condon factors are relatively diagonal, the ac Stark shifts due to each electronic state is dominated by only a few vibrational states.

\subsection{$B^2\Sigma$ ac Stark Shifts}
\label{sec:Bstark}
To estimate the $B^2\Sigma$ ac Stark shifts due to the tweezer light, we take into account coupling to the $X^2\Sigma$, $A^2\Pi_{1/2}$, $A^2\Pi_{3/2}$, $B^2\Sigma$, $C^2\Pi_{1/2}$, and $C^2\Pi_{3/2}$ electronic states. We resolve vibrational states for the $A^2\Pi_{1/2}$, $A^2\Pi_{3/2}$, $B^2\Sigma$, and $C$ states. In addition, the spin-orbit splitting of the $C^2\Pi_{1/2}$ and $C^2\Pi_{3/2}$ is ignored except for the $C^2\Pi_{1/2}(v=3)$ and $C^2\Pi_{3/2}(v=3)$ states, which are near-resonant with the trapping wavelength. As before, spin-rotation and hyperfine splittings are ignored, and Franck-Condon factors are incorporated to calculate vibrational line contributions.

While the $D$, $E$ and $F$ states could contribute to the overall ac Stark shift, the corresponding transition dipole moments are unavailable in the literature. Since the $B-E$ and $B-F$ transitions are all blue-detuned compared to the tweezer light of 781.1\,nm, they are expected to contribute a negative ac Stark shift. At our tweezer depths, the total estimated contribution from the $B-D$, $B-E$ and $B-F$ transitions is estimated to be no more than a few MHz, much smaller than the overall differential ac Stark shifts. Since the transitions are also far detuned, we expect that the contribution is roughly constant over the wavelength range explored. Based on computed Franck-Condon factors for the $B-D$ using spectroscopic data~\cite{Kaledin1999XABCaF,Gittins1993CaFCD,Hao2019dipole}, the $B-D$ transition is expected to contribute a positive ac Stark shift.

\section{Extracting $B-C$ Transition Dipole Matrix Element}
In this section, we describe in detail how the $B-C$ transition dipole matrix element is extracted from differential ac Stark-shift measurements between the $X$ and $B$ states.

\subsection{Measurement Procedure}
As described in the main text, we measure the variation of the differential ac Stark shift between the $X^2\Sigma(v=0,N=1)$ and $B^2\Sigma(v=0,N=0)$ states as a function of the wavelength of the optical tweezer light. On the $X-B$ light at 531\,nm, four sidebands each addressing the four hyperfine levels in $X^2\Sigma(v=0,N=1)$ are created using a phase modulator~\cite{holland2021gsa}. The tweezer depth is held constant, while a 5\,\text{ms} pulse of 531\,nm light illuminates the molecules. When the light is resonant with the Stark-shifted resonance, molecules are heated out of the tweezer. We detect the number of surviving molecules $30\,\text{ms}$ following the illumination pulse. As a function of the overall detuning of the $X-B$ spectroscopy light, we observe a single loss feature. The peak loss frequency as a function of the tweezer wavelength ($\lambda$) directly measures the average differential Stark shift between the $X^2\Sigma(v=0,N=1)$ and $B^2\Sigma(v=0,N=0)$ states. We assume that the measured shift is 1/2 of the peak differential ac Stark shift.

To verify the linear dependence of the Stark shift with power, we measure the differential Stark shift versus tweezer power at $\lambda=781.1\,\text{nm}$ and the magic wavelength of 782\,nm. We find that both curves scale linearly with power \,(Fig.~\ref{fig:Starkshift}(a)). Extrapolation to zero power allows us to calibrate the frequency of $X-B$ light that corresponds to zero differential ac Stark shift.

\subsection{Extracting Transition Dipole Moment from Lineshape}
The measured differential ac Stark shifts versus trapping wavelength reveals two dispersive features that originate from the $B^2\Sigma(v=0,N=0) - C^2\Pi_{1/2} (v=3,J=1/2,3/2)$ and $B^2\Sigma(v=0,N=0) - C^2\Pi_{3/2}(v=3,J=3/2)$ transitions. We fit the data to a model of the expected ac Stark shifts with three free parameters: the overall offset, the centers of the two dispersive features, and the overall strength of the two features. Shifts from states other than the $C^2\Pi (v=3)$ states are expected to contribute an approximately constant background. This can account for excited state transitions we do not explicitly consider and/or inaccurate theoretical dipole matrix elements and Franck-Condon factors for the excited electronic states.

\begin{figure}[h]
	\includegraphics[width=1.0 \columnwidth]{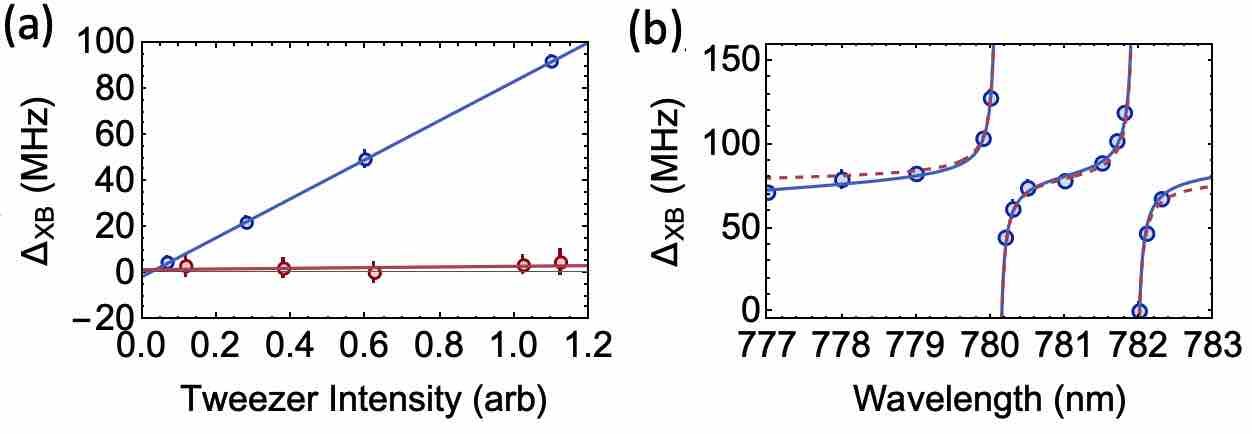}
	\caption{\label{fig:Starkshift} Measured differential ac Stark shifts $\Delta_{XB}$ between $X^2\Sigma(v=0,N=1)$ and $B^2\Sigma(v=0,N=0)$. (a) $\Delta_{XB}$ versus tweezer intensity. Shown in blue (red) circles are data for 781.1\,nm (782.0\,nm). Solid lines are linear fits to the data. The difference in the extrapolated ac Stark shift at zero intensity is consistent with the absolute frequency accuracy of $\pm 1\,\text{MHz}$ our $X$-$B$ optical frequency stabilization system. (b) $\Delta_{XB}$ versus tweezer wavelength at fixed tweezer intensity. The tweezer intensity corresponds to a trap depth of 1.28\,mK for molecules in the $X^2\Sigma(v=0,N=1)$ state. Shown in the dashed red line is a fit to the Stark shift model, where the amplitude of the $B-C$ dipole matrix element is allowed to vary and an overall offset is allowed. The fit yields a dipole moment reduction factor of 0.259(7) relative to the value in~\cite{Raouafi2001CaFdipole}, and an overall offset of 4.4(14)\,MHz. The solid blue line is a fit to the ac Stark shift model plus a term linear in the wavelength. The fit yields a dipole moment reduction factor of 0.259(5) relative to the value in~\cite{Raouafi2001CaFdipole}, and an overall offset of 5.0(10)\,MHz.}
\end{figure}

The measured Stark shift and fitted curve are shown in Fig.~\ref{fig:Starkshift}. In our fit, we take into account the calibrated tweezer depth and the known expected ac Stark shifts. The ac Stark shifts included in the Stark shift model are described in Sections~\ref{sec:Xstark} and \ref{sec:Bstark}. In brief, for the $X^2\Sigma$ state, we include ac Stark shifts from the excited $A$, $B$, $C$, $D$, $E$ and $F$ states. For the $B^2\Sigma$ state, we include Stark shifts due to the $X$, $A$, $B$, $D$ states.  To allow for slight variation in transmission and AOD diffraction efficiency as the wavelength is varied, we allow for a background term that is linear in wavelength. As shown in Fig.~\ref{fig:Starkshift}, this extended model fits the data well. We have verified that the addition of a background term that is linear in wavelength does not change the fitted constants beyond the fit errors.

From the fit, we find that the spin-orbital splitting between the $C^2\Pi_{1/2}(v=3)$ and $C^2\Pi(v=3)$ states is 280(1)GHz, agreeing well with previous spectroscopy. The extracted product of the Franck-Condon factor $f_{BC,03}$ and the dipole matrix element between $B^2\Sigma(v=0)$ and $C^2\Pi(v=3)$, $d_{BC}$, is $5.2(3) \times 10^{-3} e a_0$. Using $f_{BC,03}=2.71\times10^{-3}$ ($B^2\Sigma(v=0) - C ^2\Pi(v=3)$) computed from spectroscopy data~\cite{Kaledin1999XABCaF,Gittins1993CaFCD,Hao2019dipole}, we obtain a $B-C$ dipole moment of $2.0(1) \, e a_0$, differing significantly from the much larger theoretical value of $7.33\, e a_0$ computed in~\cite{Raouafi2001CaFdipole}. 

From the fit, the overall residual shift is 5.0(1.5)\,MHz, which is $10\%$ of the overall background offset away from the two resonances. Since most of the total Stark shift is accounted for, this provides further support that the extracted $B-C$ dipole moment is indeed much smaller than computed in~\cite{Raouafi2001CaFdipole}. If one assumes that the residual shift is entirely due to the nearby $D^2\Sigma(v=0,1,2,3,4)$ states, and uses Franck-Condon factors computed from spectroscopic data, the residual shift suggests a $B-D$ dipole moment of $1.6(3)\, e a_0$. Except for the overall offset, the uncertainties for the above are dominated by the uncertainty in the overall trap depth. We note that the uncertainties stated do not take into account uncertainties in the Franck-Condon factors.

\subsection{Comparison of Measured and Predicted Differential ac Stark Shifts}
Table~\ref{tab:expBCcontrib} lists the estimated contributions to the differential ac Stark shift to the $X^2\Sigma(v=0) - B^2\Sigma(v=0)$ transition computed with the $B-C$ dipole moment of $2.0(1) \, e a_0$. For comparison, Table~\ref{tab:theoryBCcontrib} lists the estimated contributions to the differential Stark shift with the theoretical $B-C$ transition dipole moment of $7.33\, e a_0$~\cite{Raouafi2001CaFdipole}. 

\begin{table}[h!]

\begin{center}
\begin{tabular}{ c|c }
Transition(s) & Shift (MHz) \\
\hline
$X(v=0)$ due to A,B,C,D,E,F states &  26.6 \\ 
$B(v=0)$ due to $C(v=0)$   & 27.3 \\  
$B(v=0)$ due to $C(v=1)$ & 13.3 \\
$B(v=0)$ due to $C(v=2)$ &4.1 \\
$B(v=0)$ due to $C(v=3)$ &-0.02 \\
$B(v=0)$ due to $C(v=4)$ &-0.008 \\
\hline
Total & 71.3 \\
Observed & 78(2) 
\end{tabular}
\end{center}
\caption{\label{tab:expBCcontrib} Breakdown of contributions to the $X-B$ differential ac Stark shift, using our measured value of the $B-C$ transition dipole moment $d_{BC}=2.0 ea_0$. The residual shift could be due to coupling of the $B$ state to the higher lying $D$ state.
}
\end{table}

\begin{table}[h!]
\begin{center}
\begin{tabular}{ c|c }
Transition(s) & Shift (MHz) \\
\hline
$X(v=0)$ due to A,B,C,D,E,F states &  26.6 \\ 
$B(v=0)$ due to $C(v=0)$   & 365 \\  
$B(v=0)$ due to $C(v=1)$ & 178 \\
$B(v=0)$ due to $C(v=2)$ & 54 \\
$B(v=0)$ due to $C(v=3)$ & -2.2\\
$B(v=0)$ due to $C(v=4)$ & -0.1 \\
\hline 
Total (predicted) & 621.3 \\
Observed & 78(2) 
\end{tabular}
\end{center}
\caption{\label{tab:theoryBCcontrib} Breakdown of contributions to the $X-B$ differential ac Stark shift, using our measured value of the $B-C$ transition dipole moment $d_{BC}=7.33 ea_0$ computed in~\cite{Raouafi2001CaFdipole}.}
\end{table}

\section{Parity-Conserving Rotational Loss into $X^2\Sigma(v=0, N=3)$}
As mentioned in the main text, molecules can be lost into the $X^2\Sigma(v=0,N=3)$ rotational state during laser cooling and imaging due to off-resonant photon scattering. In this section, we provide supplemental data demonstrating that this occurs, but is negligible with the addition of a $N=3$ repumper addressing the $X^2\Sigma(v=0,N=3) - B^2\Sigma(v=0,N=2)$ transition.

\begin{figure}[h!]
	\includegraphics[width=1.0 \columnwidth]{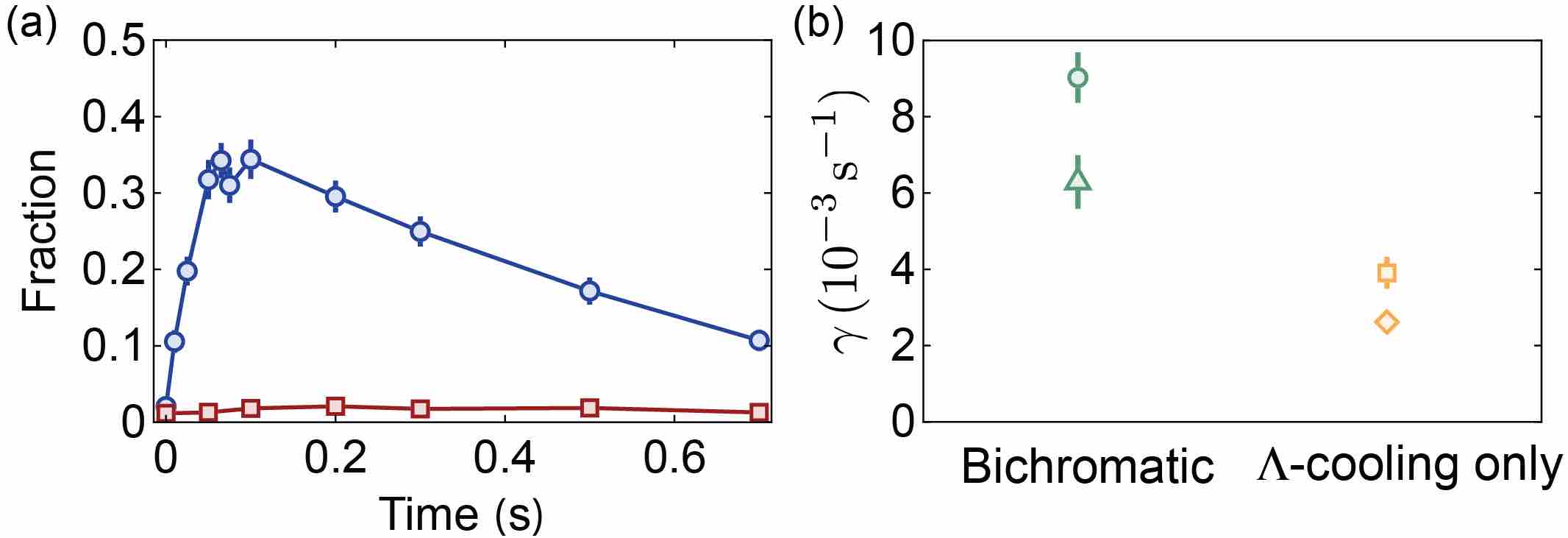}
	\vspace{-0.30in}
	\caption{\label{fig:n32} (a) Fraction of molecules in $X^2\Sigma(v=0,N=3)$ as a function of $\Lambda$-cooling time. Blue circles (red squares) show data without (with) the addition of the $N=3$ repumper. (b) Loss rate comparison. The observed loss rates for bichromatic imaging with (without) $N=3$ repumper is shown by the green triangle (circle).  The observed loss rates for $\Lambda$-cooling alone with (without) $N=3$ repumper is shown by the orange diamond (square).  }
\end{figure}

First, we examine parity-conserving rotational loss into $X^2\Sigma\,(v=0,N=3)$ with only $\Lambda$-cooling/imaging. We track the $X^2\Sigma\,(v=0,N=3)$ population as a function of $\Lambda$-imaging time. In detail, we apply $\Lambda$-imaging light that addresses the $X-A$ transition for various durations, after which we blow out all molecules in the $X^2\Sigma\,(v=0,N=1)$ using a short 2\,ms resonant pulse of  light addressing the $X^2\Sigma\,(v=0,N=1)- A^2\Pi_{1/2} (v=0, J=1/2,+)$ transition. Next, $N=3$ repumping light resonant with the $X^2\Sigma\,(v=0,N=3) - B^2\Sigma \,(v=0,N=2)$ is applied to bring molecules back into $X^2\Sigma\,(v=0,N=1)$ state for detection. As shown in Fig.~\ref{fig:n32}(a), we see a rise in $N=3$ population over about 100\,ms followed by a slower decay, as molecules are progressively off-resonantly scattered into higher rotational states $N=5,7,\ldots$. This verifies that rotational loss into $N=3$ due to off-resonant scattering of $\Lambda$-cooling light is present.

Next, we perform the same measurement sequence with the presence of the $N=3$ repumping light. In comparison, we observe no rise in $N=3$ population over the imaging timescale of 100\,ms with the experimental uncertainty, implying that this rotational leakage channel is sufficiently closed. 

Separately, we perform lifetime measurements with and without the $N=3$ repumper, both for bichromatic imaging and also for $\Lambda$-cooling alone. We find that the addition of $N=3$ repumper reduces the $\Lambda$-cooling loss rate by $1.3(4)\times10^{-3}\, \text{s}^{-1}$ from $3.9(4)\times10^{-3}\, \text{s}^{-1}$ to $2.6(1)\times10^{-3}\, \text{s}^{-1}$, and reduces the bichromatic imaging loss rate by $2.7(9)\times10^{-3}\, \text{s}^{-1}$ from $9.0(6)\times10^{-3} \text{s}^{-1}$ to $6.3(6)\times10^{-3}\, \text{s}^{-1}$. Within experimental uncertainty, we find that during bichromatic imaging, the loss into $N=3$ can be attributed almost entirely to the $\Lambda$-cooling light. This is expected, since during bichromatic imaging, the intensity of the $X-A$ cooling light is higher than the $X-B$ imaging light ($240\,\text{mW/cm}^2$ versus $90\,\text{mW/cm}^2$), and the relevant excited state that gives rise to $N=3$ leakage is further detuned for the $X-B$ transition compared to that of the $X-A$ $\Lambda$-cooling transition (60\,GHz versus 30\,GHz). We note in passing that because the $X(v=0)-A(v=0)$ and $X(v=0)-B(v=0)$ transitions are highly diagonal, the decay rate into the higher vibrational states $X^2\Sigma \, (v=1,2,3,\ldots, N=3)$ is suppressed by a factor of $\approx 80$ and $\approx 500$, implying that these processes are much slower than the relevant experiment timescales.

\section{Rate Equation Model for Modeling Population Dynamics During Imaging}
We extract the parity-changing loss rate during bichromatic imaging through fitting population dynamics measurements to a rate equation model. In this section, we describe in detail the rate equation model, and how the parameters of the model are determined.  

To enhance the effect of parity-changing loss when the molecules are illuminated by $X$-$B$ light, we make all population dynamic measurements at a higher $X-B$ intensity of $24(2)\,\text{mW/cm}^2$, which is double the optimal bichromatic imaging $X-B$ intensity. For detection, the $X-B$ intensity is kept at the optimal value of $12(1)\,\text{mW/cm}^2$.

 \subsection{Relevant States and Processes}
In our rate equation model, the relevant states described are $\left|X,v=0,N=0, J=1/2, F=1, m_F=-1,0,1\ket$, $\left|X,v=0,N=0,J=1/2, F=0, m_F=0\ket$, $\left|X,v=0,N=1\ket$, and $\left|X,v=0,N=2\ket$. These are denoted by $N_{0,\left|1,m_F\ket}$, $N_{0,\left|0,0\ket}$,$N_{1}$, and $N_{2}$, respectively. Note that we keep track of the populations in the $\left|X,v=0,N=0, J=1/2\ket$ hyperfine $m_F$ states separately, since some of our measurements either prepare or detect molecules in only one of these states.

During bichromatic imaging,  due to photon scattering, population can be transferred between or lost from the above states. Our model captures the following population transfer and loss processes. 

First, bichromatic imaging empirically results in loss of molecules from the detectable $X^2\Sigma (v=0, N=1)$ manifold. We denote the total imaging loss rate by $\Gamma_{\text{img}}$, which we directly measure. This loss rate includes all possible loss channels out of $X^2\Sigma (v=0, N=1)$, known and unknown. 

Second, we separately include parity-changing terms that arise from molecules decaying via the $B\rightarrow A \rightarrow X$ pathway. During bichromatic imaging, molecules are excited on the $X^2\Sigma(v=0, N=1, J=1/2, F=0) - B^2\Sigma(v=0, N=0, J=1/2)$ transition. The excited molecules can subsequently decay either directly to $X^2\Sigma$ or via the two-photon pathways $B^2\Sigma \rightarrow A^2\Pi_\Omega \rightarrow X$, where $\Omega=1/2,3/2$. In the latter processes, the parity of the molecules is changed upon returning to $X^2\Sigma$, and the molecules end up in $X^2\Sigma(N=0)$ or $X^2\Sigma(N=2)$. We describe these in detail in Section~\ref{sec:xabbranch}.

Third, during imaging, off-resonant photon scattering also leads to population transfer between various even-parity ground rotational states $X^2\Sigma(v=0,N=0,2)$. The dominant process is $\Lambda$-cooling light that is resonant with the $X^2\Sigma(v=0,N=1) \rightarrow A^2\Pi_{1/2}(v=0,J=1/2,+)$ transition, off-resonantly addressing the $X^2\Sigma(v=0,N=0,2)\rightarrow A^2\Pi_{1/2}(v=0,J=1/2,3/2,5/2,7/2,-)$. We take into account off-resonant transitions to states in the $A^2\Pi_{1/2}(v=0)$ manifold. We will describe these in detail in Section~\ref{sec:offrestrans} and Section~\ref{sec:offresbranch}. 

There are three other much slower processes that occur, which are not included in the rate equation model. These three processes are as follows. First, Raman scattering of the tweezer light can lead to rotational mixing of the $X^2\Sigma(v=0,N=0)$ and $X^2\Sigma(v=0,N=2)$ states, and also further loss into higher rotational states with even $N$. We have measured that this occur on timescales much longer than the explored internal dynamics here. Second, loss to $X^2\Sigma(v=1)$ can also occur due to blackbody radiation, but is expected to occur on a much longer timescale (observed to be $\sim 5\,\text{s}$). Third, losses due to collision with background gas leads to state-independent loss. Lifetime measurements of $X^2\Sigma(N=1)$ molecules indicate that this occurs on a much longer timescale even when compared to blackbody excitation.

\subsubsection{$X-A-B$ Branching Ratios}
\label{sec:xabbranch}
\begin{figure}
	\includegraphics[width=1.0 \columnwidth]{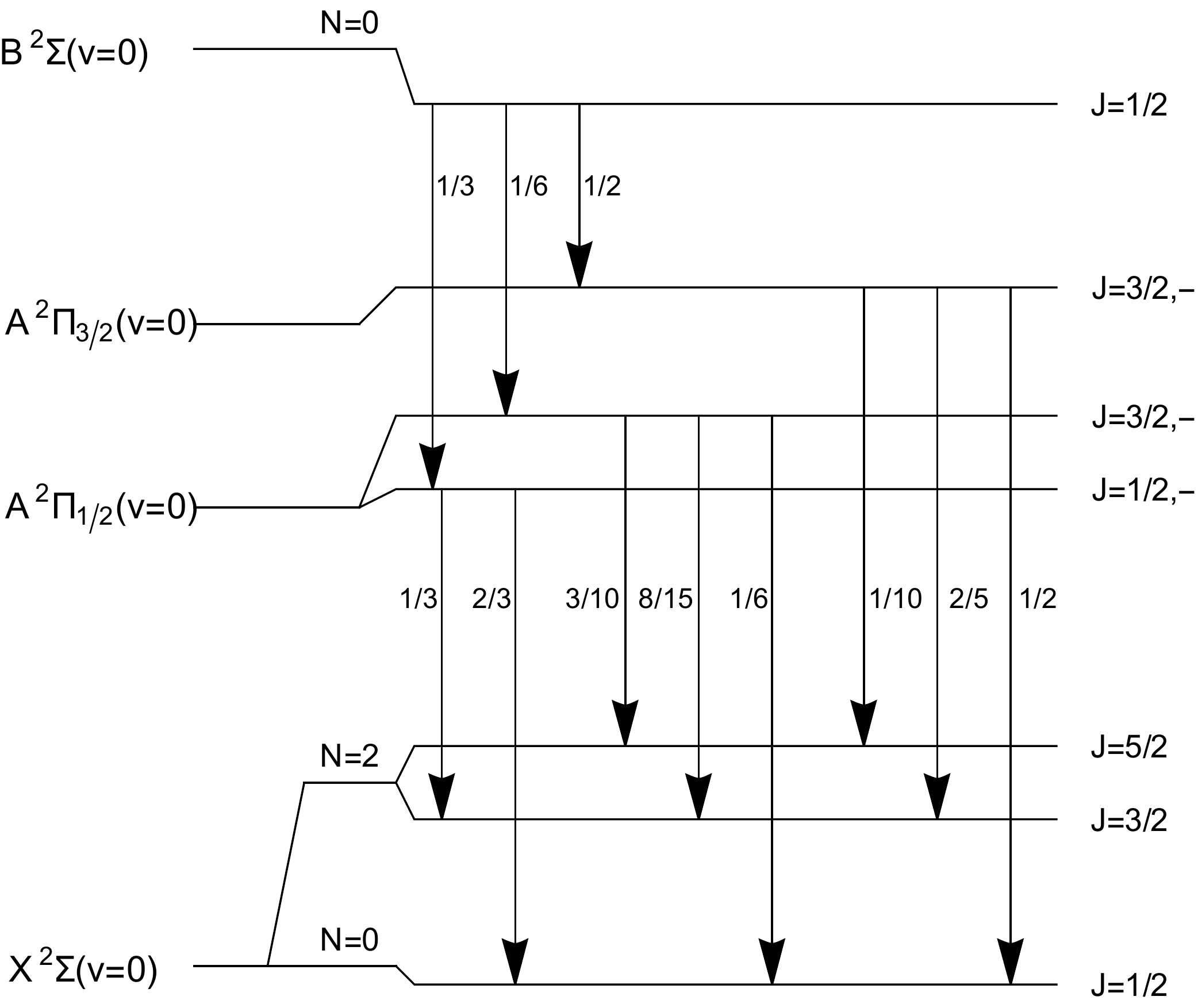}
	\caption{\label{fig:BAXbranch} $B\rightarrow A \rightarrow X$ branching ratios.}
\end{figure}

We first compute the parity-changing branching ratios of excited molecules in $B^2\Sigma(v=0,N=0)$ into $A^2\Pi_{\Omega = 1/2,3/2}(v=0,J)$ manifolds, and the subsequent branching ratios into the $X^2\Sigma(v=0,N=0,2,J)$ manifolds. We take into account rotation and spin-rotation coupling but ignore hyperfine structure, which is small in the excited $A$ and $B$ states. The relevant branching ratios are shown in Fig.~\ref{fig:BAXbranch}. We make the approximation that the $B^2\Sigma(v=0) \rightarrow A^2\Pi_{1/2}(v=0)$ and $B^2\Sigma(v=0) \rightarrow A^2\Pi_{3/2}(v=0)$  transitions frequencies are identical.

\subsubsection{Off-Resonant Scattering Processes}
\label{sec:offrestrans}
The nominally dark $X^2\Sigma(v=0, N=0,2)$ manifolds experience off-resonant scattering from the bichromatic imaging light. Predominantly, the bichromatic imaging light consists of $\Lambda$-cooling light addressing the $X^2\Sigma(v=0, N=1)\rightarrow A^2\Pi_{1/2}(v=0,J=1/2,+)$ transition and vibrational repumping light addressing the $X^2\Sigma(v=1, N=1)\rightarrow A^2\Pi_{1/2}(v=0,J=1/2,+)$ transition. The $X-A$ $\Lambda$-cooling light is detuned by $10 - 100\,\text{GHz}$ from transitions  addressing the $X^2\Sigma(v=0, N=0,2)$ manifolds. On the other hand, the $v=1$ vibrational repumping light is detuned by $\sim 17\,\text{THz}$. Its effect is therefore negligible in comparison and is not taken into account.

\begin{figure}
	\includegraphics[width=1.0 \columnwidth]{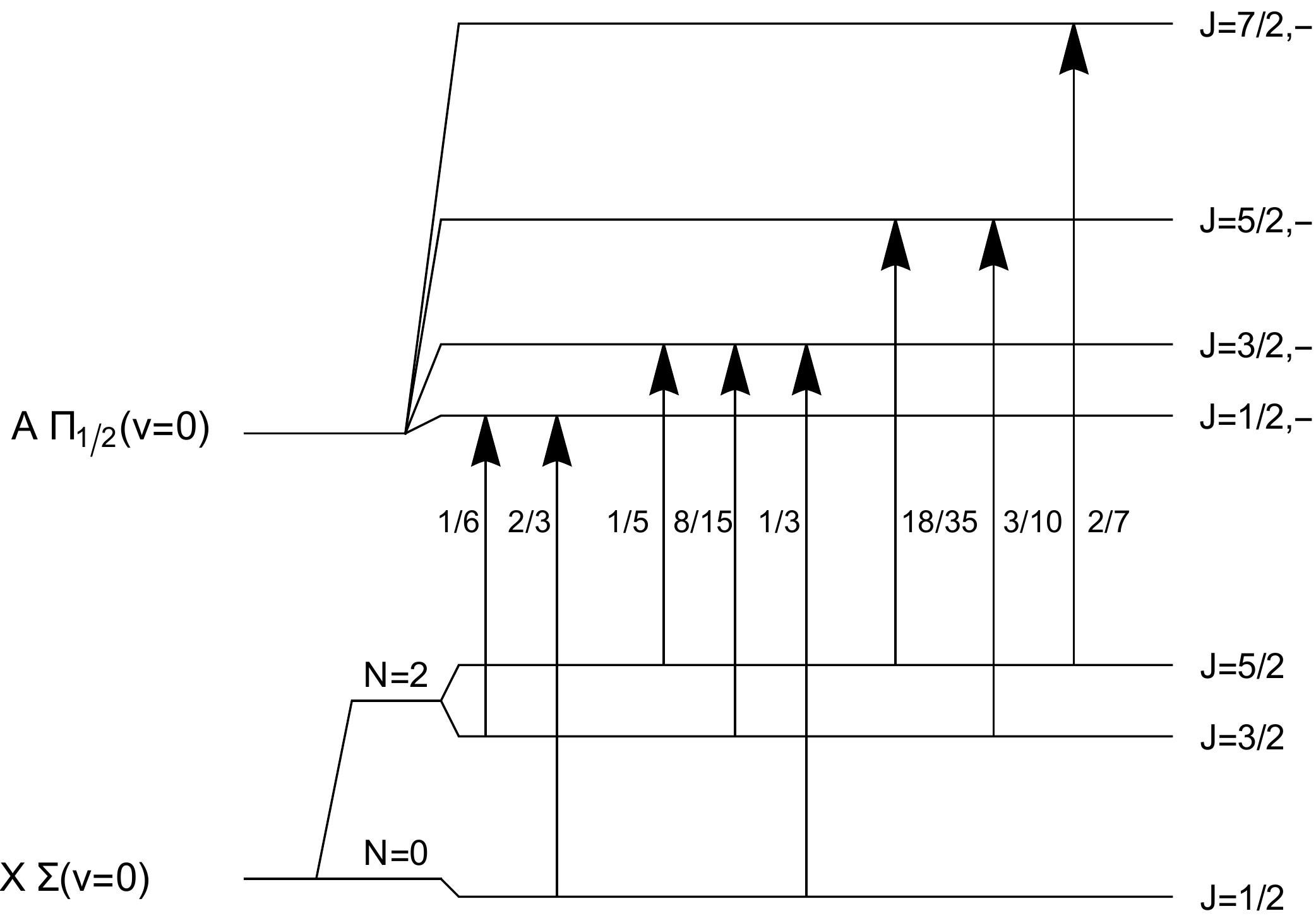}
	\caption{\label{fig:HonlLondon.pdf} Relative H{\"o}nl-London factors in the $A^2\Pi_{1/2}(v=0,-)-X^2\Sigma(v=0,+)$ system. The factors are polarization averaged and summed over all final states.}
\end{figure}

The off-resonant scattering on the $X^2\Sigma(v=0, N=0,2)$ manifolds arise from transitions to the $A^2\Pi_{1/2}(v=0, J=1/2,3/2,5/2,7/2)$ excited manifolds. The off-resonant scattering rate due to each transition is determined by H{\"o}nl-London factors (Fig.~\ref{fig:HonlLondon.pdf}), and its detuning $\Delta_i$ from the $X-A$ cooling light.

In Table~\ref{tab:detunings}, we list the relevant transitions, their detuning $\Delta_i$ from the $X-A$ $\Lambda$-cooling light, and the relative H{\"o}nl-London factors $s_i$. Since all the detunings are large compared to the excited $A$-state linewidth of $\Gamma_A = 2\pi \times 8\,\text{MHz}$, the photon scattering rates are given by$\frac{2I d_{XA}^2}{\varepsilon_0 c} \frac{s_i}{3\Delta_i^2}\Gamma_A$, where $I$ is the intensity of the $X-A$ $\Lambda$-cooling light and $d_{XA}$ is the reduced electric transition dipole moment between $X$ and $A$ electronic states. The H{\"o}nl-London factors $s_i$ average over all light polarizations and sum over excited Zeeman ($m_J$) states. We will ignore hyperfine coupling in both $X$ and $A$ states.

\begin{table*}
\begin{center}
\begin{tabular}{ c |c |c | c }
 Transition & Detuning $\Delta_i$ & Strength $s_i$ & Rate\\ 
\hline
$X^2\Sigma(v=0,N=0,J=1/2) \rightarrow A^2\Pi_{1/2}(v=0,J=1/2,-) $& $-19.6\,\text{GHz}$ &2/3 & $\Gamma_{0l}$\\  
$ X^2\Sigma(v=0,N=0,J=1/2) \rightarrow A^2\Pi_{1/2}(v=0,J=3/2,-) $& $-52.6\,\text{GHz}$  & 1/3 & $\Gamma_{0h}$\\  
\hline 
$  X^2\Sigma(v=0,N=2,J=3/2) \rightarrow A^2\Pi_{1/2}(v=0,J=1/2,-) $& $+42.4\,\text{GHz}$ &1/6 &$ \Gamma_{2l0}$\\   
$  X^2\Sigma(v=0,N=2,J=3/2) \rightarrow A^2\Pi_{1/2}(v=0,J=3/2,-) $& $+9.0\,\text{GHz}$  & 8/15& $ \Gamma_{2l1}$\\      
 $ X^2\Sigma(v=0,N=2,J=3/2) \rightarrow A^2\Pi_{1/2}(v=0,J=5/2,-) $& $-39.8\,\text{GHz}$  & 3/10& $\Gamma_{2l2}$\\   
 \hline
$ X^2\Sigma(v=0,N=2,J=5/2) \rightarrow A^2\Pi_{1/2}(v=0,J=3/2,-) $&  $+9.0\,\text{GHz}$ & 1/5 & $\Gamma_{2h1}$\\    
 $ X^2\Sigma(v=0,N=2,J=5/2) \rightarrow A^2\Pi_{1/2}(v=0,J=5/2,-)$ & $-39.8\,\text{GHz}$  &18/35& $\Gamma_{2h2}$\\    
  $      X^2\Sigma(v=0,N=2,J=5/2) \rightarrow A^2\Pi_{1/2}(v=0,J=7/2,-) $& $-117.8\,\text{GHz}$  &2/7& $\Gamma_{2h3}$\\    
\end{tabular}
\end{center}
\caption{\label{tab:detunings} Transitions in the even parity sector of $X^2\Sigma(v=0)$ relevant for two-photon scattering. The respective detunings away from the $\Lambda$-cooling light on the $X^2\Sigma(v=0,N=1) - A^2\Pi_{1/2}(v=0,N=0,J=1/2,+)$ transitions are listed. Here we ignore $X$-state spin-rotation and hyperfine splittings, and ignore $A$-state hyperfine coupling. Detunings are relative to line center.}
\end{table*}

Off-resonant scattering leads to remixing of $X^2\Sigma(v=0,N=0)$ and $X^2\Sigma(v=0,N=2)$ populations, vibrational loss, and also loss to same-parity $X^2\Sigma(v=0)$ states with higher $N$. 

\subsubsection{Branching Ratios for Even-Parity States due to Off-Resonant Scattering}
\label{sec:offresbranch}
\begin{figure}[t]
	\includegraphics[width=1.0 \columnwidth]{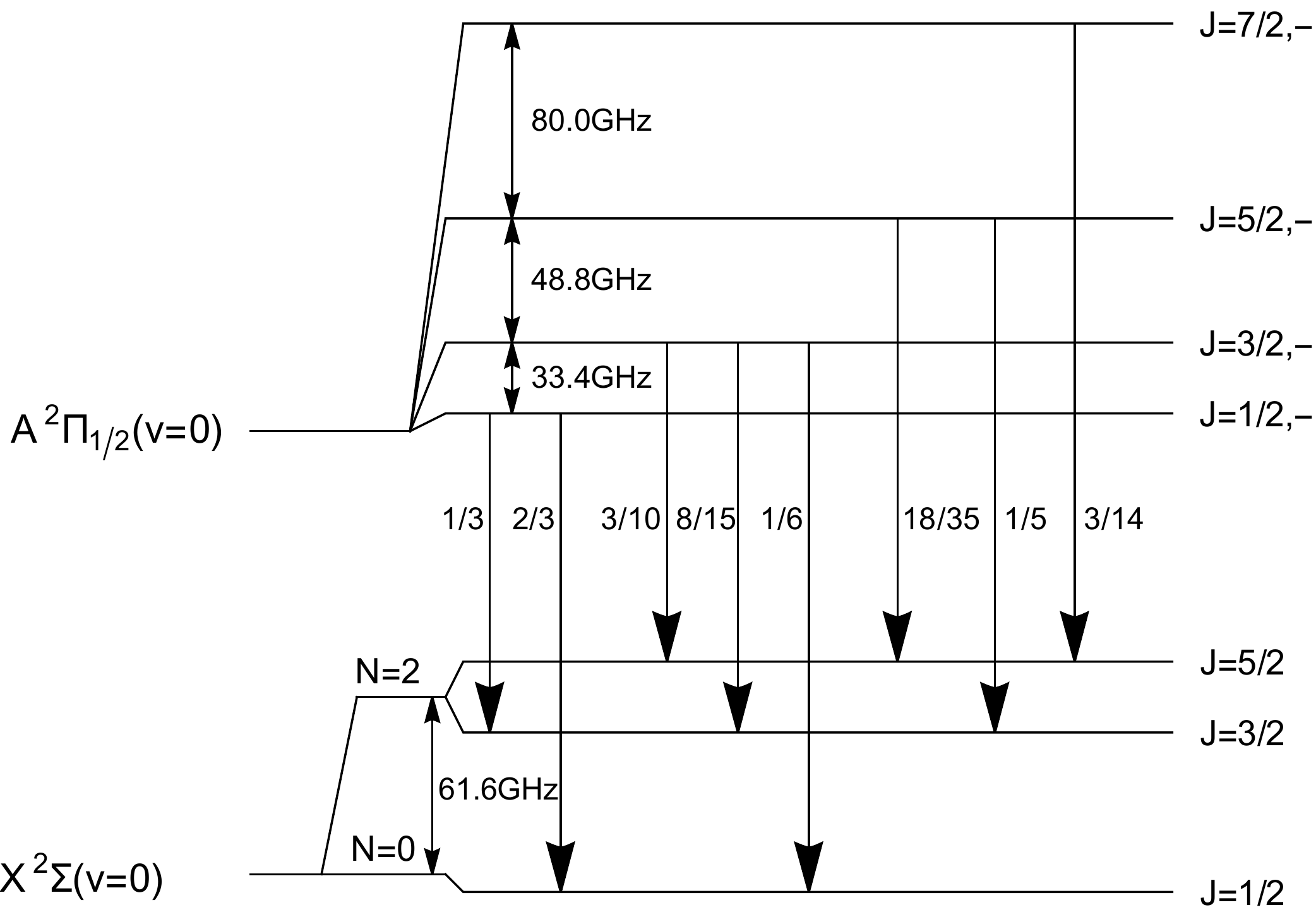}
	\caption{\label{fig:AXbranch} Relevant off-resonant scattering transitions, energy level spacings, and branching ratios in the $A^2\Pi_{1/2}(v=0,-)-X^2\Sigma(v=0,+)$ system. }
\end{figure}

Averaged over all hyperfine states, a molecule excited into $A^2\Pi_{1/2}(v=0,J=1/2,3/2,-)$ can decay into $X^2\Sigma(v=0, N=0)$ and $X^2\Sigma(v=0, N=2)$, while a molecule excited into $A^2\Pi_{1/2}(v=0,J=5/2,7/2,-)$ can decay into $X^2\Sigma(v=0, N=2)$ and  $X^2\Sigma(v=0, N=4)$. The decay probabilities are shown in Fig.~\ref{fig:AXbranch}.

\subsection{Rate Equation Model}
We describe the population dynamics under illumination of bichromatic imaging light ($\Lambda$-cooling light on $X^2\Sigma(v=0, N=0) \rightarrow A^2\Pi_{1/2}(v=0,J=1/2,-)$, along with imaging light on the $X^2\Sigma(v=0, N=1, J=1/2, F=0) \rightarrow B^2\Sigma(v=0, N=0)$) with the following rate equation model: 
\begin{widetext}
\begin{eqnarray}
\dot{N}_{0,\left|1,1\ket} &=&- (\Gamma_{0l}+\Gamma_{0h}) N_{0,\left|1,1\ket} +\frac{1}{4}  f_{XA}\left(\frac{2}{3}\Gamma_{0l}+\frac{1}{6}\Gamma_{0h}\right)   N_0 \nonumber \\
& & + \frac{1}{4} f_{XA} \left(\frac{2}{3} \Gamma_{2l0}+\frac{1}{6} \Gamma_{2l1} \right)N_{2,3/2} + \frac{1}{4} f_{XA} \left(\frac{1}{6} \Gamma_{2h1} \right)N_{2,5/2} +\frac{1}{4} \frac{1}{2} f_{\text{BXA}} \Gamma_{\text{parity}} N_1 \nonumber \\
\dot{N}_{0,\left|1,0\ket} &=&- (\Gamma_{0l}+\Gamma_{0h})  N_{0,\left|1,0\ket} +\frac{1}{4}  f_{XA} \left(\frac{2}{3}\Gamma_{0l}+\frac{1}{6}\Gamma_{0h}\right)   N_0\nonumber \\
& & + \frac{1}{4} f_{XA} \left(\frac{2}{3} \Gamma_{2l0}+\frac{1}{6} \Gamma_{2l1} \right)N_{2,3/2} + \frac{1}{4} f_{XA} \left(\frac{1}{6} \Gamma_{2h1} \right)N_{2,5/2} +\frac{1}{4} \frac{1}{2} f_{\text{BXA}} \Gamma_{\text{parity}} N_1 \nonumber \\
\dot{N}_{0,\left|1,-1\ket} &=&- (\Gamma_{0l}+\Gamma_{0h})  N_{0,\left|1,-1\ket} +\frac{1}{4}  f_{XA} \left(\frac{2}{3}\Gamma_{0l}+\frac{1}{6}\Gamma_{0h}\right)   N_0 \nonumber \\
& & + \frac{1}{4} f_{XA} \left(\frac{2}{3} \Gamma_{2l0}+\frac{1}{6} \Gamma_{2l1} \right)N_{2,3/2} + \frac{1}{4} f_{XA} \left(\frac{1}{6} \Gamma_{2h1} \right)N_{2,5/2} +\frac{1}{4} \frac{1}{2} f_{\text{BXA}} \Gamma_{\text{parity}} N_1 \nonumber \\
\dot{N}_{0,\left|0,0\ket} &=&-\frac{3}{4} \frac{1}{2} \Gamma_0 N_{0,\left|0,0\ket} +\frac{1}{4}  f_{XA} \left(\frac{2}{3}\Gamma_{0l}+\frac{1}{6}\Gamma_{0h}\right)   N_0\nonumber \\
& & + \frac{1}{4} f_{XA} \left(\frac{2}{3} \Gamma_{2l0}+\frac{1}{6} \Gamma_{2l1} \right)N_{2,3/2} + \frac{1}{4} f_{XA} \left(\frac{1}{6} \Gamma_{2h1} \right)N_{2,5/2} +\frac{1}{4} \frac{1}{2} f_{\text{BXA}} \Gamma_{\text{parity}} N_1 \nonumber \\
\dot{N}_{1} &=&-\Gamma_{\text{parity}} N_1 - (\Gamma_{\text{img}}-\Gamma_{\text{parity}}) N_1 \nonumber \\
\dot{N}_{2,3/2} &=& -( \Gamma_{2l0}+\Gamma_{2l1}+\Gamma_{2l2} )N_{2,3/2}  + f_{\text{XA}} \left(\frac{1}{3}\Gamma_{0l}+\frac{8}{15}  \Gamma_{0h}\right)N_0 + f_{\text{XA}} \left(\frac{1}{3}\Gamma_{2l0}+\frac{8}{15}  \Gamma_{2l1}+\frac{1}{5}  \Gamma_{2l2}\right)N_{2,3/2}\nonumber \\
& & + f_{\text{XA}} \left(\frac{8}{15}  \Gamma_{2h1}+\frac{1}{5}  \Gamma_{2h2}\right)N_{2,5/2}+ \frac{2}{5} f_{\text{BXA}} \Gamma_{\text{parity}} N_1 \nonumber \\
\dot{N}_{2,5/2} &=& -( \Gamma_{2h1}+\Gamma_{2h2}+\Gamma_{2h3} )N_{2,5/2}+ f_{\text{XA}}\frac{3}{10}\Gamma_{0h} N_0  + f_{\text{XA}} \left(\frac{3}{10}  \Gamma_{2l1}+\frac{18}{35}  \Gamma_{2l2}\right)N_{2,3/2}\nonumber \\
& & + f_{\text{XA}} \left(\frac{3}{10}  \Gamma_{2h1}+\frac{18}{35}  \Gamma_{2h2}+\frac{3}{14}  \Gamma_{2h3}\right)N_{2,5/2} + \frac{1}{10}  f_{\text{BXA}} \Gamma_{\text{parity}} N_1, \nonumber
\end{eqnarray}
\end{widetext}
where $N_0 = N_{0,\left|1,1\ket}+ N_{0,\left|1,0\ket}+ N_{0,\left|1,1\ket}+N_{0,\left|1,-1\ket}+N_{0,\left|0,0\ket}$, and $f_{XA}$ is the $v=0 \rightarrow v'=0$ Franck-Condon factor for the $A^2\Pi_{1/2} \rightarrow X^2\Sigma$ system.
The transitions corresponding to the various photon scattering rates $\Gamma_i$ can be found using Table.~\ref{tab:detunings}.

\subsection{Determining Rate Equation Parameters}
\subsubsection{Determining Overall Imaging Loss Rate}
To make progress, we first measure the overall imaging loss rate $\Gamma_{\text{img}}$ at the $X-B$ intensity used for all population measurements. This intensity is twice that used for the detection images, in order to enhance the relative contribution of parity-changing loss. The imaging loss rate is obtained by measuring the remaining $X^2\Sigma (v=0, N=1)$ population versus imaging time. By fitting the data to an exponentially decaying curve, we obtain $\Gamma_{\text{img}} = 14.1(16)\,\text{s}^{-1}$ (Fig.~\ref{fig:rateequations}(a)). 

\subsubsection{Determining Off-Resonant Scattering Rate}
We next determine the various off-resonant scattering rates. Since the off-resonant scattering rates are all related by their respective detunings and Honl-London factors, we can express all off-resonant scattering rates to $\Gamma_{0l}$.

\begin{figure}[h]
	\includegraphics[width=1.0\columnwidth]{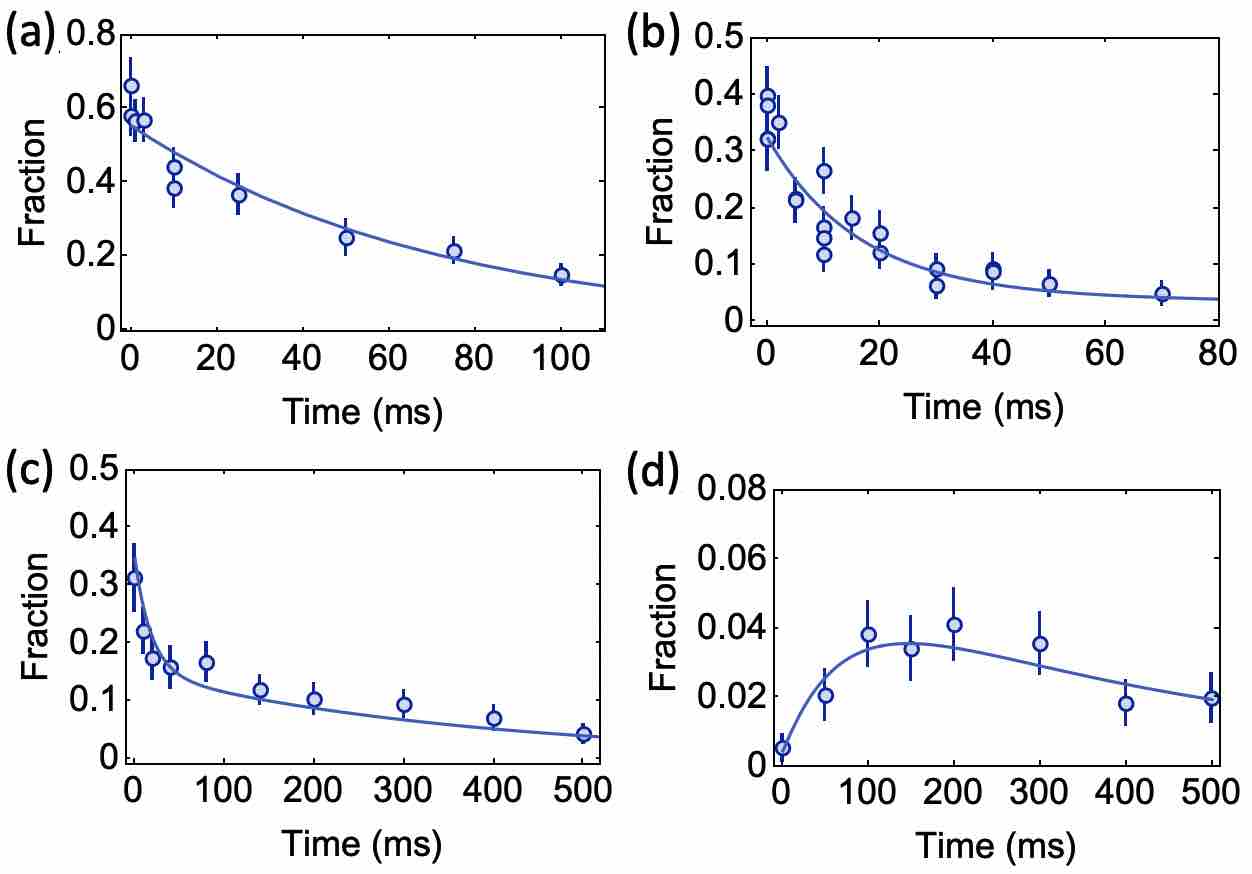} 
	\caption{\label{fig:rateequations} (a) $X^2\Sigma (v=0, N=1)$ population versus bichromatic imaging time. Solid line shows a fit to an exponential decay curve with no vertical offset. (b) $\left| X(v=0), N=0, J=1/2, F=1,m_F=1\ket$ population versus bichromatic imaging time. Solid line shows the result of a simultaneous fit to our rate equation model. (c) Total $\left| X(v=0), N=0, J=1/2, F=1,m_F\ket$ population versus bichromatic imaging time. Solid line shows the result of a simultaneous fit to our rate equation model. For (b,c), the fit parameters are two overall scale factors and the off-resonant scattering rate. (d) Total $\left| X(v=0), N=0, J=1/2, F=1,m_F\ket$ population versus bichromatic imaging time, with molecules initially in the $X^2\Sigma(v=0, N=1)$ manifold. Solid line shows the result of a fit to our rate equation model. 
	}
\end{figure}

To determine $\Gamma_{0l}$, we first prepare molecules in $\left| X, v=0, N=0, F=1, m_F=1\ket$. Next, we illuminate them with bichromatic imaging light for variable durations, before detecting the remaining population in $\left| X, v=0, N=0, F=1, m_F=1\ket$. In detail, molecules loaded into the tweezers are first imaged with bichromatic light for 10\,ms to determine their initial occupation. Subsequently, an optical pumping pulse pumps a fraction of molecules into $\left| X, v=0, N=1, J=1/2, F=0\ket$. We then perform a Landau-Zener microwave frequency sweep at a magnetic field of 4.3\,G to transfer $\left| X, v=0, N=1, J=1/2, F=0\ket$ molecules into $\left| X, v=0, N=0, J=1/2, F=1, m_F=1\ket$. Subsequently, a short (2\,ms) pulse of light resonant with $X^2\Sigma (v=0, N=1)\rightarrow A^2\Pi(v=0, J=1/2,+)$ removes all remaining population in the $X^2\Sigma(v=0, N=1)$ manifold. To detect the remaining population in $\left| X, v=0, N=0, F=1, m_F=1\ket$, a reverse microwave sweep on the $\left| X, v=0, N=1,J=1/2, F=0\ket \rightarrow  \left| X, v=0, N=0, J=1/2, F=1,m_F=1\ket$ is applied to bring molecules back into the $X^2\Sigma (v=0, N=1)$ manifold, which is subsequently detected with a 30\,ms long imaging pulse. As shown in Fig.~\ref{fig:rateequations}(b), under bichromatic imaging, the $\left|X,v=0,N=0,J=1/2,F=1,m_F=1\ket$ population decays rapidly over $20\,\text{ms}$. 

To provide further information on the population dynamics, we perform a second measurement, where population in all $\left| X, v=0, N=0, F=1, m_f\ket$ states are detected. In detail, after the variable bichromatic imaging pulse, three microwave sweeps are performed sequentially on the $\left| X, v=0, N=0,J=1/2, F=1,m_F=-1,0,1\ket \rightarrow \left| X, v=0, N=1,J=1/2, F=0\ket $ transitions, and interspersed with a short ``depumping'' pulse ($50\,\mu\text{s}$) resonant with the $\left| X, v=0, N=1,J=1/2, F=0\ket \rightarrow \left| A^2\Pi_{1/2} , v=0, J=1/2,+\ket$ transition. The depumping pulse optically pumps molecules out of the $X(v=0,N=1,F=0)$ state into the other $X(v=0,N=1)$ states. As shown in Fig.~\ref{fig:rateequations}(c), in this second measurement, we observe a rapid decay over $\sim 20\,\text{ms}$, followed by a long decay over $\sim 300\,\text{ms}$.

The above two curves are then fit simultaneously to a solution of the rate equation model with initial population entirely in $N_{0,\left|1,1\ket}$. Note that $N_1$, $\Gamma_{\text{img}}$ and $\Gamma_{\text{parity}}$ are irrelevant since no population is in $X(v=0,N=1)$. To fit the two curves, we allow separate amplitude scaling factors to take into account imperfect initial state preparation and detection that might differ between the two measurements. The two curves are fit with a single off-resonant scattering rate $\Gamma_{0l}$, since all the other loss rates except $\Gamma_{\text{img}}$ and $\Gamma_{\text{parity}}$ can be expressed in terms of $\Gamma_{0l}$. We find that the data is described well by the rate equation model (Fig.~\ref{fig:rateequations}(b,c)), yielding an experimentally measured off-resonant photon scattering rate $\Gamma_{0l}=59(7)\,\text{s}^{-1}$. Furthermore, the fitted amplitudes allow us to determine that at long times, the second measurement sequence detects $3.2(4)$ times more molecules than the first sequence. This is consistent with the ideal factor of 3, which results from measuring three $X(v=0,N=0,F=1)$ states instead of just a single one. The extracted amplitude scaling factors also agree, providing additional evidence that the rate equation model does describe the population dynamics well.

As a cross-check, we estimate $\Gamma_{0l}$ using the known beam powers. During $\Lambda$-imaging, six 10\,mm beams with a total of $\sim 120\,\text{mW}$ of $X^2\Sigma(v=0,N=1) \rightarrow A^2\Pi_{1/2}(v=0,J=1/2,+)$ light addresses the molecules. Using the Honl-London factors and detunings, this provides a predicted scattering rate of $\Gamma_{0l,\text{pred}} = 60(20)\,\text{s}^{-1}$, where the uncertainty is estimated from the uncertainty in the intensity. The agreement provides further confidence in our model.

\subsection{Extracting the Parity-Changing Rate and $B-A$ Branching Ratio}

Having determined $\Gamma_{\text{img}}$ and all the off-resonant scattering rates, we next determine the parity-changing rate $\Gamma_{\text{parity}}$. We prepare molecules in the  $X^2\Sigma(v=0, N=1)$ manifold and perform bichromatic imaging for a variable amount of time. Next, $X^2\Sigma(v=0, N=1)$ molecules are removed using a short resonant pulse of imaging light ($2\,\text{ms}$). We next transfer molecules from the $\left|X(v=0), N=0, F=1,m_F\ket$ states into the $X^2\Sigma(v=0, N=1)$ manifold via microwave sweeps interspersed with depumping pulses (described in previous section). The final $X^2\Sigma(v=0, N=1)$ population is measured. 

As shown in Fig.~\ref{fig:rateequations}(d), the detected population is initially zero. As the bichromatic imaging duration increases ($X-B$ intensity still at double the optimal bichromatic imaging value), we observe a rise over $\sim 100\,\text{ms}$ followed by a slower decay. In order to fit to the rate equation model, we need to determine the overall scale, i.e. what fraction of initially identified molecules are in fact available to use in the measurement, and what fraction of remaining molecules are detected. 

We separately measure overall survival efficiencies and state transfer efficiencies. Based on the initial detected fraction in the $N=1$ lifetime measurement, we estimate that 0.55(3) of the tweezers initially identified as occupied in the first image are subsequently detected. This takes into account loss during the first image, classification errors in the first and second images, and any other loss between the two images that occurs due to technical heating and collisions with background gas. The lower fraction than the optimal non-destructive detection fidelity is in part due to classification errors in the second image, and also an intentionally higher classification threshold used in the second image. The higher threshold is used to minimize false positives, which could lead to an offset. Second, we measure the efficiency of the microwave sweep used to bring molecules into $N=1$ for detection. By tracking the decay in contrast after a series of Landau-Zener sweeps, we find a transfer efficiency of 0.84(6).

Taking into account the uncertainty in the overall scale, and fitting the resulting dynamics of the detected population with $\Gamma_{\text{img}}$ and the off-resonant scattering rates fixed at the previously determined values, we determine $\Gamma_{\text{parity}} = 3.3(6)\,\text{s}^{-1}$. Here, the error bar is the statistical fitting error with all other parameters fixed. The fit has as a free parameter a small offset, which takes into account imperfect removal of $X^2\Sigma(v=0,N=1)$ molecules during the detection sequence.

Assuming that the parity loss is entirely due to the $ B\rightarrow A \rightarrow X$ channel, and using the separately measured $X \rightarrow B$ photon scattering rate, we obtain a $B\rightarrow A$ branching ratio of $8.4(15)\times 10^{-5}$. This is much higher than the previously measured $N=0$ and $N=2$ branching rates of $7(1)\times 10^{-6}$ and $1.6(3) \times 10^{-6}$~\cite{Truppe2017chirp}, but about $35\%$ lower than the theoretically predicted branching ratio of $1.3 \times 10^{-4}$~\cite{Hao2019dipole}. Converting to a transition dipole matrix element, this gives an $A-B$ transition dipole moment of $0.30(3)\,e a_0$, $0.07\, e a_0$ smaller than computed~\cite{Hao2019dipole}.

\section{Relating AC Stark Shifts to Loss Rates}
\label{sec:starktoloss}
In this section, we describe how ac Stark shifts can be related to photon scattering loss rates. In our system, the close detuning of optical tweezer trapping light to the $B-C$ transitions leads to relatively high admixture of the $C^2\Pi $ states into the $B^2\Sigma \,(v=0)$ state, which molecules occupy during bichromatic imaging. In detail, inside the tweezer trap, the $B^2\Sigma \,(v=0)$, which we denote by $\left | B'\ket$, acquires an admixture $\beta_i$ of a $C^2\Pi $ state denoted by $\left |C_i\ket$. Denoting the dressed state as $\left | B'\ket$, one finds that
\begin{equation}
\left |B' \ket \approx \left| B\ket + \sum_i \beta_i \left |C_i\ket.
\end{equation}
Because the $C^2\Pi $ state has non-diagonal Franck-Condon factors with the lower-lying $X$, $A$ and $B$ states, and because cascaded decays with multiple photons are possible when a molecule decays through intermediate states into the $X^2\Sigma$ state, a $C^2\Pi $ molecule that decays will remain in the $X^2\Sigma (v=0,1,2,3, N=1)$ optical cycling states with low probability. Making the approximation that a molecule excited into $C^2\Pi $ states is always lost from optical cycling, one finds that the dressed $\left |B'\ket$ state acquires an additional loss rate $\gamma_{ad}$ of 
\begin{eqnarray}
\gamma_{ad} &=& \sum_i \gamma_i \nonumber \\
\gamma_i &=& \beta_i^2 \Gamma_i, 
\end{eqnarray}
where $\Gamma_i$ is the inverse lifetime of $\left |C_i\ket$. 

Assuming that the lifetime of the $C^2\Pi$ states are only determined by E1 radiative decays, sum rules state that all the $\Gamma_i$ are approximately identical. We therefore set $\Gamma_i = \Gamma_{C, \text{rad}}$, the E1 radiative decay rate of 
of the $C$ state, which can be estimated from the transition dipole moments to the lower-lying $X$, $A$, and $B$ states.

To make further progress, we need to determine $\beta_i$. This can be determined from the ac Stark shifts. To simplify the discussion, we next provide relations between ac Stark shifts and scattering rates for a two-level system under the rotating wave approximation.  

We consider a two-level system with states $\left| g \right \rangle$ and  $\left| e\right  \rangle$, with energy spacing of $\hbar \omega_0$. The system is then placed in the presence of off-resonant light with electric field $E$ oscillating at an angular frequency of $\omega$. In the rotating wave approximation, the system is described by the following Hamiltonian:
\begin{equation}
\hat{H} = \hbar \left(\begin{array}{ccc} \delta/2 & \Omega/2 \\ \Omega/2 & -\delta/2 \end{array}\right),
\end{equation}
where $\delta = \omega_0-\omega$, and $\Omega = E d$ where $d$ is the transition dipole moment between $\left | g\ket$ and $\left |e \ket$.  

In the limit of $\delta \gg \Omega$, we can apply perturbation theory to find that the dressed ground state becomes
\begin{eqnarray}
\left |g'  \ket & \approx & \left |g \ket  + \beta \left |e \ket, \nonumber \\
\beta &=& \frac{\Omega}{2\delta}
\end{eqnarray}
and acquires an energy shift of 
\begin{equation}
\Delta = -\hbar \frac{\Omega^2}{4\delta}
\end{equation}

Denoting the decay rate of $\left |e\ket$ as $\Gamma$, one finds that the dressed ground state has a photon scattering rate $\Gamma_{sc}$ of 
\begin{equation}
\Gamma_{sc} = \frac{\Delta}{\hbar \delta} \Gamma.
\end{equation}

For the specific case in the text, the rotating wave approximation is well-justified for the $B-C$ transitions. The additional loss rate due to $C$-state admixture into the dressed $B$ state $\left|B'\ket$ can then be written as
\begin{equation}
\gamma_{ad} = \Gamma_c \sum_i \frac{\Delta_i}{\hbar \delta_i},
\end{equation}
where $\Delta_i$ is the ac Stark shift due to the $i^{\text{th}}$ $C^2\Pi$ state, and $\delta_i = \omega_{0,i} - \omega$, where $\omega_{0,i}$ is the resonance angular frequency for the $i^{\text{th}}$ $B-C$ transition. Note that $\Delta_i$ is a function of the tweezer wavelength.

Lastly, during imaging, the dressed state $\left |B'\ket$ is addressed by $X-B$ imaging light nominally addressing the $X^2\Sigma \,(v=0, N=1) - B^2\Sigma \, (v=0,N=0)$ transition. Since only a fraction of molecules $P_B$ are in the $\left|B'\ket$ state, the additional loss rate due to $C$-state admixture during bichromatic imaging is given by
\begin{eqnarray}
\gamma_{b,ad} &=& P_B \gamma_{ad} \nonumber \\
P_B &=& \frac{\Gamma_{B,sc}}{\Gamma_{B}},
\end{eqnarray}
where $\Gamma_{B,sc}$ is the photon emission rate of $X-B$ photons, and $\Gamma_{B}$ is the linewidth of the $B$ state. We directly measure $\Gamma_{B,sc}$, and use a previously measured value for  $\Gamma_{B}$~\cite{Dagdigian1974XB}. We note that in our loss estimates, we include contributions from the $C^2\Pi_{\Omega=1/2,3/2} (v=0,1,2,3,4)$ states, but only contributions from the two $C^2\Pi_{\Omega=1/2,3/2} (v=3)$ states are significant.

\section{Loss Rate Accounting versus Tweezer Wavelength}
\begin{figure}
	\includegraphics[width=1.0 \columnwidth]{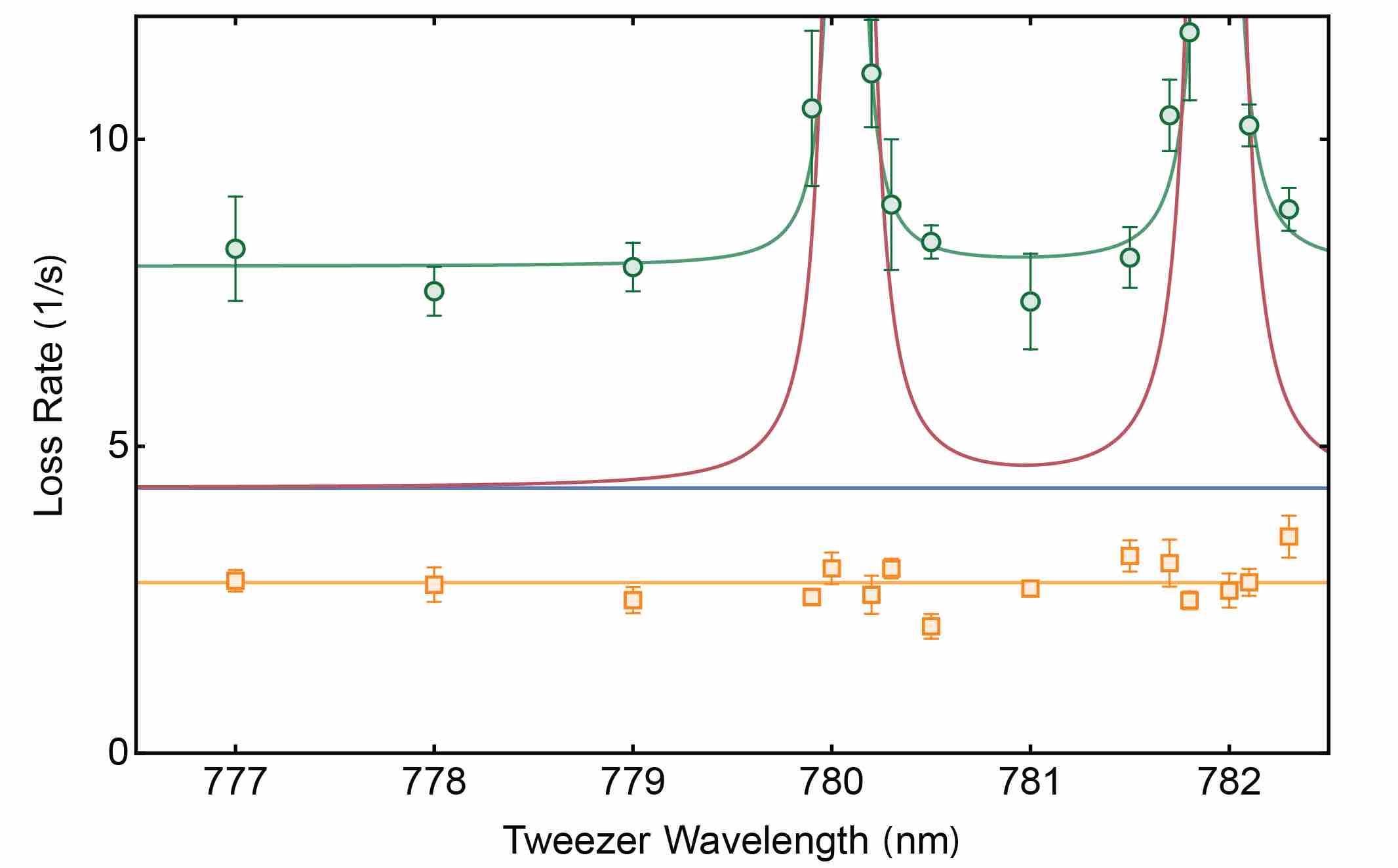}
	\caption{\label{fig:allloss}Loss Rate versus Tweezer Wavelength. Shown in green circles is the bichromatic imaging loss rate $\gamma$, at the $X-B$ imaging intensity used for detection. The green line shows a fit to two Lorentzians with an offset. Orange squares show the loss rate from $\Lambda$-cooling $\gamma_\Lambda$ versus tweezer wavelength; no dependence is observed. The orange line shows the average value. The blue line shows the estimated loss due to $\Lambda$-cooling and parity-changing loss from $B\rightarrow A \rightarrow X$ decays using our measured value of $d_{AB}$. The red line shows the estimated loss due to $\Lambda$-cooling, parity-changing loss from $B\rightarrow A \rightarrow X$, and loss due to admixture of the $C$ state into the $B$ state.}
\end{figure}

In this section, we provide supplemental information on the imaging loss rate versus tweezer wavelength. In addition to bichromatic imaging loss rate $\gamma$, we also measure the  $\Lambda$-imaging loss rate $\gamma_{\Lambda}$. As described in the main text, the difference between these two rates is the excess loss rate $\gamma_{\text{excess}} = \gamma - \gamma_{\Lambda}$, which is shown in Fig.~3(a) and Fig.~4(h) in the main text. In Fig.~\ref{fig:allloss}, we show in addition, $\gamma$ and $\gamma_{\Lambda}$ versus tweezer wavelength, and the estimated loss from $C$-state admixture and $X-A-B$ decays using our measured values of $d_{AB}$ and $d_{BC}$, and theoretical values of $d_{BA}$ and $d_{BX}$~\cite{Raouafi2001CaFdipole}. As shown in Fig.~\ref{fig:allloss}, the loss due to $\Lambda$-cooling alone shows no dependence on the trapping wavelength over the explored range.

\section{Estimation of the $C^2\Pi$ Linewidth}
We estimate the $C^2\Pi$ linewidth by assuming that it is completely determined by E1 radiative decays to the lower-lying $X$, $A$, and $B$ states. From the extracted $B-C$ transition dipole moment, and the theoretical values of the $A-C$ and $X-C$ transition dipole moments from~\cite{Raouafi2001CaFdipole}, we estimate the $C$-state linewidth to be $\Gamma_C = 2\pi \times 5.2(2) \,\text{MHz}$. This is in contrast to a linewidth of $\Gamma_C^{th} = 2\pi \times 29 \,\text{MHz}$ obtained from using the $B-C$ transition dipole moment of $7.33\,e a_0$ computed in~\cite{Raouafi2001CaFdipole}.

As a cross-check, we also separately estimate the $C$-state linewidth through the dependence of excess loss $\gamma_{\text{excess}}$ on tweezer wavelength. Using the results derived in Section~\ref{sec:starktoloss}, we fit the loss data versus wavelength to
\begin{equation}
\gamma_{\text{excess}} = \gamma_{bg} + \gamma_{b,ad}, 
\end{equation} 
with $\Gamma_C$ as a free parameter, and $\gamma_{bg}$ allowing a non-zero offset. The model fits the data well (Fig.~\ref{fig:allloss}), and provides $\Gamma_C = 2\pi \times 3.9(6)\,\text{MHz}$, which is in rough agreement with the estimated E1 radiative decay rate $\Gamma_{C,\text{rad}}=2\pi \times 5.3(2)\,\text{MHz}$ obtained using the measured value of $d_{BC}$. Since this method could have systematic effects that are difficult to quantify, we only use this to rule out the possibility that $\Gamma_{C}$ is much larger.

\bibliographystyle{apsrev4-1}